\documentclass[12pt]{article}
\usepackage[margin=1in]{geometry}
\usepackage{amsmath}
\usepackage{amsfonts}
\usepackage{graphicx,psfrag,epsf}
\usepackage{enumerate}
\usepackage{natbib}
\usepackage{url} 
\usepackage{prodint}
\usepackage{amsthm}
\usepackage{multirow}
\usepackage{booktabs}

\newtheorem{thm}{Theorem}

\newcommand{\blind}{1}
\newcommand{\GG}[1]{}

\begin{document}

\def\spacingset#1{\renewcommand{\baselinestretch}%
{#1}\small\normalsize} \spacingset{1}


\if1\blind
{
  \title{\bf Semiparametric Marginal Regression for Clustered Competing Risks Data with Missing Cause of Failure}
  \author{Wenxian Zhou\hspace{.2cm}\\
    Department of Biostatistics, Indiana University\\
    Giorgos Bakoyannis\hspace{.2cm}\\
    Department of Biostatistics, Indiana University\\
    Ying Zhang\hspace{.2cm}\\
    Department of Biostatistics, University of Nebraska Medical Center\\
    Constantin T. Yiannoutsos\hspace{.2cm}\\
    Department of Biostatistics, Indiana University\\
    }
  \maketitle
} \fi

\if0\blind
{
  \bigskip
  \bigskip
  \bigskip
  \begin{center}
    {\LARGE\bf Semiparametric Marginal Regression for Clustered Competing Risks Data with Missing Cause of Failure}
\end{center}
  \medskip
} \fi

\bigskip
\spacingset{2} 
\begin{abstract}
Clustered competing risks data are commonly encountered in multicenter studies. The analysis of such data is often complicated due to informative cluster size, a situation where the outcomes under study are associated with the size of the cluster. In addition, cause of failure is frequently incompletely observed in real-world settings. To the best of our knowledge, there is no methodology for population-averaged analysis with clustered competing risks data with informative cluster size and missing causes of failure. To address this problem, we consider the semiparametric marginal proportional cause-specific hazards model and propose a maximum partial pseudolikelihood estimator under a missing at random assumption. To make the latter assumption more plausible in practice, we allow for auxiliary variables that may be related to the probability of missingness. The proposed method does not impose assumptions regarding the within-cluster dependence and allows for informative cluster size. The asymptotic properties of the proposed estimators for both regression coefficients and infinite-dimensional parameters, such as the marginal cumulative incidence functions, are rigorously established. Simulation studies show that the proposed method performs well and that methods that ignore the within-cluster dependence and the informative cluster size lead to invalid inferences. The proposed method is applied to competing risks data from a large multicenter HIV study in sub-Saharan Africa where a significant portion of causes of failure is missing.
\end{abstract}

\noindent%
{\it Keywords:} Clustered data; Competing risks; Informative cluster size; Missing cause of failure.
\vfill

\section{Introduction}
\label{intro}
Clustered competing risks data are commonly encountered in multicenter studies \citep{Zhou12, Diao13}. An important feature of such data is that the outcomes of individuals from the same cluster are typically dependent and, thus, the standard assumption of independence is violated \citep{Balan19, Bakoyannis19Biometrics}. Therefore, standard methods for competing risks data are not applicable in the presence of clustering.
In addition, cluster size is often informative, in the sense that the outcomes under study are associated with cluster size \citep{Pavlou13, Seaman14, Cong07}. An example of a setting with informative cluster size is a study about dental health outcomes. In such studies, dental health outcomes are related to the total number of teeth (cluster size) in a given person \citep{Williamson08}. With informative cluster size, the standard methods for clustered data lead to bias, since larger clusters are over-represented in the sample and have a larger influence on the parameter estimates. In addition, cause of failure is frequently incompletely observed in real-world settings due to nonresponse/missingness or by the study design \citep{Bakoyannis20}. A complete case analysis which discards observations with missing event types is expected to lead to bias and efficiency loss \citep{Lu01, Bakoyannis10}.

This work is motivated by a large multicenter HIV cohort study conducted by the East Africa Regional Consortium of the International Epidemiology Databases to Evaluate AIDS (EA-IeDEA). A major goal of the study was to evaluate HIV healthcare clinics in East Africa and study two important outcomes in HIV care: (i) disengagement from care and (ii) death while in care (i.e. prior to disengagement). 
In this study, patients who received antiretroviral treatment (ART) at the same clinic are expected to have correlated outcomes. In addition, the number of patients in the clinic is expected to be associated with the outcomes of interest, since clinics with more patients are typically better staffed and are expected to provide better care. Furthermore, there was a severe death under-reporting issue in sub-Saharan Africa, which implies that a patient who was lost to clinic was either deceased (and the death was not reported to the clinic) or had disengaged from care. To deal with this issue, EA-IeDEA implemented a double-sampling design where a small subset of patients who were lost to clinic were intensively outreached in the community and their vital status was actively ascertained. This double-sampling design transformed the misclassification problem (i.e. unreported deaths) into a missing data problem by the study design, where cause of failure (i.e. death while in care or disengagement) was unknown for the non-outreached lost patients.

There are two main classes of models for dealing with the within-cluster dependence issue for survival data. One is frailty models \citep{Clayton85, Hougaard86, Liu11}, which specify explicitly the within-cluster dependence via random effects and provide cluster-specific inference. Such models typically impose assumptions about the structure of the within-cluster dependence and the distribution of the random effects, and tend to be computationally intensive. Under this class of models, \citet{Katsahian06} proposed a frailty proportional hazards model for the subdistribution of a competing risk, while \citet{Scheike10} proposed a semiparametric random effects model for competing risks data where the interest is on a particular cause of failure. The other class of models is marginal models \citep{Wei89, Liang93, Cai97, Spiekerman98, Cai00}. These models do not rely on assumptions regarding the dependence structure and have a population-averaged interpretation. Following this idea, \citet{Zhou12} proposed a marginal version of the Fine-Gray model \citep{Fine99} for population-averaged analysis of clustered competing risks. The issue of informative cluster size with survival outcomes has been addressed via a within-cluster resampling method \citep{Cong07} and a weighted score function approach \citep{Williamson08}, where the weights are equal to the inverse of the number of observations in the corresponding cluster.

The issue of missing cause of failure with independent competing risks data has received considerable attention in the literature \citep{Goetghebeur95, Lu01, Craiu04, Gao05, Lu08, Bakoyannis10, Hyun12, Bordes14, Nevo17, Bakoyannis20}. Recently, \citet{Bakoyannis20} proposed a unified framework for semiparametric regression and risk prediction for competing risks data with missing at random (MAR) cause of failure, under the proportional cause-specific hazards model. Unlike previous methods, the approach by \citet{Bakoyannis20} provides inference for both regression and functional parameters such as the cumulative incidence function. The latter quantity is key for risk prediction in modern medicine. Moreover, simulation studies have shown that the approach by \citet{Bakoyannis20} provides substantially more efficient regression parameter estimates compared to augmented inverse probability weighting estimators \citep{Gao05, Hyun12} and the multiple-imputation estimator \citep{Lu01}. However, all the aforementioned methods did not consider a potential within-cluster dependence and are thus expected to lead to invalid inferences with clustered data. To the best of our knowledge, only \citet{Lee17} have addressed the issue of analyzing clustered competing risks data with missing cause of failure. \citet{Lee17} proposed a frailty proportional cause-specific hazards model along with a hierarchical likelihood approach for estimation. Nevertheless, this approach does not allow for informative cluster size, it imposes strong assumptions regarding the within-cluster dependence and the distribution of the frailty, which may be violated in practice, and does not provide inference for the infinite-dimensional parameters such as the cumulative incidence function. In addition, the method provides cluster-specific inference and not population-averaged inference which is more scientifically relevant in many applications, including our motivating multicenter HIV study.

To the best of our knowledge, there is no general method for population-averaged inference based on clustered competing risks data with informative cluster size and missing causes of failure. To address this problem, we consider the semiparametric marginal proportional cause-specific hazards model and propose a maximum partial pseudolikelihood estimator under a MAR assumption. The proposed method does not impose assumptions regarding the within-cluster dependence and allows for informative cluster size. Moreover, the method can be easily implemented using off-the-self software that allows for case weights, such as the function \texttt{coxph} in the R package \texttt{survival} (details regarding computation using R are provided in Web Appendix A). The proposed estimators are shown to be strongly consistent and asymptotically normal. Closed-form variance estimators are provided and rigorous methodology for the calculation of simultaneous confidence bands for the infinite-dimensional parameters is proposed. Simulation studies show that the method performs well and that the previously proposed method for missing causes of failure by \citet{Bakoyannis20}, which ignores the within-cluster dependence and the potential informative cluster size, leads to invalid inferences. Finally, the method is applied to the data from the EA-IeDEA study for illustration. 

\section{Methodology}
\label{s:method}

\subsection{Notation and Assumptions}
\label{ss:notation}
Let $i=1,2, \ldots, n$ index the $n$ clusters in the study and $j=1,2, \ldots, M_{i}$ index the subjects in the $i$th cluster. Also, let $T_{i j}$ and $U_{i j}$ denote the failure and right censoring times for the $j$th subject in the $i$th cluster. The corresponding observed counterparts are the minimum of the event or censoring times $X_{i j}=T_{i j} \wedge U_{i j}$ and the failure (from any cause) indicator $\Delta_{i j}=I(T_{i j} \leq U_{i j})$. Here we consider the finite observation interval $[0, \tau]$, for an arbitrary $\tau<\infty$. Suppose that there are $k$ competing causes of failure, with $k<\infty$, and let $C_{i j} \in\{1,2, \ldots, k\}$ denote the cause of failure for the $j$th subject in the $i$th cluster. For the sake of generality, cluster size $M$ is assumed to be random and informative, in the sense that there is an association between the event time and/or cause of failure and $M$. However, our proposed methodology applies trivially to simpler situations with non-informative or fixed cluster size. To incorporate missingness in the cause of failure, we define the missing indicator $R_{i j}$, with $R_{i j}=1$ indicating that the cause of failure for the $j$th subject in the $i$th cluster is observed, and $R_{i j}=0$ otherwise. As in previous works on missing cause of failure \citep{Bakoyannis20}, we consider the situation where right censoring status is always observed, that is if $\Delta_{i j}=0$ then $R_{i j}=1$. The cause of failure $C_{i j}$ is only observed when both $\Delta_{i j}=1$ and $R_{i j}=1$. Let $\epsilon_{i j}=\Delta_{i j} R_{i j} C_{i j}$ be the observed cause of failure, with $\epsilon_{i j}=0$ denoting the cause of failure is missing or censored. The vector of covariates of scientific interest is denoted by $\boldsymbol{Z}_{i j}\in\mathbb{R}^p$. In addition, let $\boldsymbol{A}_{i j} \in \mathbb{R}^{q}$ denote a vector of auxiliary variables, which may not be of scientific interest, but may be related to the probability of missingness. It has been argued that such auxiliary covariates can be used to make the MAR assumption more plausible in practice \citep{Lu01,Nevo17,Bakoyannis19, Bakoyannis20}. As usual, $(T_{ij}, C_{ij})$ and $U_{ij}$ are assumed independent given $\boldsymbol{Z}_{ij}$. In addition, $(T_{ij}, C_{ij}, U_{ij})$ are assumed independent across clusters conditionally on $\boldsymbol{Z}_{ij}$. However, within cluster $i$, $(T_{i j}, C_{i j})$, $j=1, \ldots, M_{i}$, are allowed to be dependent given $\boldsymbol{Z}_{ij}$, with an arbitrary dependence structure. Similarly, the right censoring times $U_{i j}$ may be dependent within cluster $i$. The observed data are $n$ i.i.d. copies of $\boldsymbol{D}_{i}=(\boldsymbol{X}_{i}, \boldsymbol{\Delta}_{i}, \boldsymbol{\epsilon}_{i}, \boldsymbol{Z}_{i}, \boldsymbol{A}_{i}, \boldsymbol{R}_{i}, M_{i})$, $i=1, \ldots, n$, where $\boldsymbol{X}_{i}=\{X_{i j}\}_{j=1}^{M_{i}}$, $\boldsymbol{\Delta}_{i}=\{\Delta_{i j}\}_{j=1}^{M_{i}}$, $\boldsymbol{\epsilon}_{i}=\{\epsilon_{i j}\}_{j=1}^{M_{i}}$, $\boldsymbol{Z}_{i}=\{\boldsymbol{Z}_{i j}\}_{j=1}^{M_{i}}$, $\boldsymbol{A}_{i}=\{\boldsymbol{A}_{i j}\}_{j=1}^{M_{i}}$, and $\boldsymbol{R}_{i}=\{R_{i j}\}_{j=1}^{M_{i}}$. To facilitate the presentation of the proposed estimator and its properties, we define the  counting process $N_{i j}(t)=I(X_{i j} \leq t, \Delta_{i j}=1)$ and at-risk process $Y_{i j}(t)=I(X_{i j} \geq t)$. Additionally, we define the cause-specific counting process as $N_{i j l}(t)=I(X_{i j} \leq t, \Delta_{i j l}=1)=\Delta_{i j l} N_{i j}(t)$, where $\Delta_{i j l}=I(C_{i j}=l, \Delta_{i j}=1)$, for $l=1, \ldots k$.

Letting $\boldsymbol{W}_{i j}=(X_{ij}, \boldsymbol{Z}_{i j}, \boldsymbol{A}_{i j})$, we impose the MAR assumption $P(R_{i j}=1|C_{i j}, \Delta_{i j}=1, \boldsymbol{W}_{i j})=P(R_{i j}=1|\Delta_{i j}=1, \boldsymbol{W}_{i j})$. This assumption is equivalent to
\begin{eqnarray*}
P(C_{i j}=l|R_{i j}=1,\Delta_{i j}=1, \boldsymbol{W}_{i j})&=& P(C_{i j}=l|R_{i j}=0,\Delta_{i j}=1, \boldsymbol{W}_{i j})\\
&=& P(C_{i j}=l|\Delta_{i j}=1,\boldsymbol{W}_{i j}) \\
&\equiv&\pi_l(\boldsymbol{W}_{i j},\boldsymbol{\gamma}_{0}), \quad l=1, \ldots, k,
\end{eqnarray*}
where $\pi_l(\boldsymbol{W}_{i j},\boldsymbol{\gamma}_{0})$ is the marginal probability of the failure cause $l$ given $\boldsymbol{W}_{i j}$, for a non-right-censored observation, and $\boldsymbol{\gamma}_{0}$ is assumed to be a finite-dimensional parameter. In Section 2.3 we provide a goodness-of-fit approach for evaluating the appropriateness of this model in practice.

\subsection{Estimation Approach}

In this work, we provide estimators and inference methodology for both marginal cause-specific hazards and cumulative incidence functions. The covariate-specific marginal cause-specific hazards are defined as
\[
\lambda_{l}\left(t ; \boldsymbol{z}\right)=\lim_{h \rightarrow 0} \frac{P\left(t \leq T_{i j}<t+h, C_{i j}=l \mid T_{i j} \geq t, \boldsymbol{Z}_{i j}=\boldsymbol{z}\right)}{h}, \quad l=1, \ldots k,
\]
and the covariate-specific marginal cumulative incidence functions are defined as
\begin{eqnarray}
F_{l}\left(t ; \boldsymbol{z}\right)&=&P\left(T_{i j} \leq t, C_{i j}=k \mid \boldsymbol{Z}_{i j}=\boldsymbol{z}\right)\nonumber\\
&=&\int_{0}^{t} \exp \left\{-\sum_{l=1}^{k} \Lambda_{l}\left(s ; \boldsymbol{z}\right)\right\} \lambda_{l}\left(s ; \boldsymbol{z}\right) ds, \quad l=1, \ldots k, \label{cif}
\end{eqnarray}
where $\Lambda_{l}(t ; \boldsymbol{z})=\int_{0}^{t} \lambda_{l}(s ; \boldsymbol{z}) ds$, which is the covariate-specific cumulative hazard for the $l$th cause of failure. Here, we adopt the marginal proportional cause-specific hazards model
\begin{equation*}
\lambda_{l}\left(t ; \boldsymbol{z}\right)=\lambda_{0,l}(t) \exp \left(\boldsymbol{\beta}_{0,l}^{T} \boldsymbol{z}\right), \quad l=1, \ldots, k,
\end{equation*}
where $\lambda_{0,l}(t)$ is the $l$th unspecified baseline cause-specific hazards function.

When there are no missing causes of failure (i.e. $R_{ij}=1$ for $j=1,\ldots,M_i$ and $i=1,\ldots,n$), estimation for clustered competing risks data can be performed, under the working independence assumption,
using the logarithm of the weighted partial likelihood for $\boldsymbol{\beta}=(\boldsymbol{\beta}_{1}, \ldots, \boldsymbol{\beta}_{k})$
\begin{eqnarray}
pl_{n}(\boldsymbol{\beta})=\sum_{l=1}^{k} \sum_{i=1}^{n} \frac{1}{M_{i}} \sum_{j=1}^{M_{i}} \int_{0}^{\tau}\left[\boldsymbol{\beta}_{l}^{T} \boldsymbol{Z}_{i j}-\log \left\{\sum_{p=1}^{n} \frac{1}{M_{p}} \sum_{q=1}^{M_{p}} Y_{p q}(t) \exp \left(\boldsymbol{\beta}_{l}^{T} \boldsymbol{Z}_{p q}\right)\right\}\right] d N_{i j l}(t). \nonumber\\
\label{loglik}
\end{eqnarray}
This can be seen as the competing risks analogue of the weighted log-partial likelihood by \citet{Cong07}, where the contribution of each subject is weighted by the inverse of the corresponding cluster size to account for informative cluster size.

When cause of failure is missing for some individuals, the weighted log-partial likelihood \eqref{loglik} cannot be evaluated for the observations with a missing cause of failure. For such situations, we propose a weighted partial pseudolikelihood estimator for $\boldsymbol{\beta}$ which replaces the unobserved cause-specific counting processes with their conditional expectation given the observed data $\boldsymbol{D}_{ij}$. These conditional expectations are equal to
\[
\tilde{N}_{i j l}\left(t ; \boldsymbol{\gamma}_{0}\right)\equiv E\{N_{i j l}(t)|\boldsymbol{D}_{ij}\}=\left\{R_{i j} \Delta_{i j l}+\left(1-R_{i j}\right) \pi_{l}\left(\boldsymbol{W}_{i j}, \boldsymbol{\gamma}_{0}\right)\right\} N_{i j}(t),
\]
The resulting logarithm of the expected partial pseudolikelihood conditional on the observed data $\{\boldsymbol{D}_{i j}\}_{i=1, \ldots n ; j=1, \ldots M_{i}}$ is
\begin{eqnarray}
Q_n(\boldsymbol{\beta})= \sum_{l=1}^{k} \sum_{i=1}^{n} \frac{1}{M_{i}} \sum_{j=1}^{M_{i}} \int_{0}^{\tau}\left[\boldsymbol{\beta}_{l}^{T} \boldsymbol{Z}_{i j}-\log \left\{\sum_{p=1}^{n} \frac{1}{M_{p}} \sum_{q=1}^{M_{p}} Y_{p q}(t) \exp \left(\boldsymbol{\beta}_{l}^{T} \boldsymbol{Z}_{p q}\right)\right\}\right] d \tilde{N}_{i j l}\left(t ; \boldsymbol{\gamma}_{0}\right).\nonumber\\
\label{Eloglik}
\end{eqnarray}
The unknown parameter $\boldsymbol{\gamma}_{0}$ in \eqref{Eloglik} needs to be replaced with an estimate $\hat{\boldsymbol{\gamma}}_n$. Such an estimate can be obtained by fitting the marginal binary or multinomial logistic model on the complete cases using generalized estimating equations and under a working independence assumption. Then, estimation of $\boldsymbol{\beta}_0$ can be performed using the partial pseudoscore function
\[
\boldsymbol{G}_{n,l}\left(\boldsymbol{\beta}; \hat{\boldsymbol{\gamma}}_n\right)=\frac{1}{n} \sum_{i=1}^{n} \frac{1}{M_{i}} \sum_{j=1}^{M_{i}} \int_{0}^{\tau}\left\{\boldsymbol{Z}_{i j}-\boldsymbol{E}_{n}\left(t, \boldsymbol{\beta}_{l}\right)\right\} d \tilde{N}_{i j l}\left(t ; \hat{\boldsymbol{\gamma}}_n\right), \quad l=1, \ldots, k,
\]
where
\[
\boldsymbol{E}_{n}\left(t, \boldsymbol{\beta}_{l}\right)=\frac{\sum_{p=1}^{n} \frac{1}{M_{p}} \sum_{q=1}^{M_{p}} Y_{p q}(t) \exp (\boldsymbol{\beta}_{l}^{T} \boldsymbol{Z}_{p q}) \boldsymbol{Z}_{p q}}{\sum_{p=1}^{n} \frac{1}{M_{p}} \sum_{q=1}^{M_{p}} Y_{p q}(t) \exp (\boldsymbol{\beta}_{l}^{T} \boldsymbol{Z}_{p q})}.
\]
The estimators $\hat{\boldsymbol{\beta}}_{n,l}$ are the solutions to $\boldsymbol{G}_{n,l}(\hat{\boldsymbol{\beta}}_{n,l},\hat{\boldsymbol{\gamma}}_n)=\boldsymbol{0}$, $l=1, \ldots, k$. This estimation procedure can be easily implemented using \texttt{coxph} function in the R package \texttt{survival} with some data management. An illustration of the use of the \texttt{coxph} function to obtain parameter estimates with the proposed approach is provided in Web Appendix A.

For $l=1, \ldots, k$ and $t \in [0, \tau]$, the Breslow-type estimator for the marginal cause-specific baseline cumulative hazard function is
\[
\hat{\Lambda}_{n,l}(t)=\sum_{i=1}^{n} \frac{1}{M_{i}} \sum_{j=1}^{M_{i}} \int_{0}^{t} \frac{d \tilde{N}_{i j l}(u ; \hat{\boldsymbol{\gamma}}_n)}{\sum_{p=1}^{n} \frac{1}{M_{p}} \sum_{q=1}^{M_{p}} Y_{p q}(u) \exp (\hat{\boldsymbol{\beta}}_{n,l}^{T} \boldsymbol{Z}_{p q})}.
\]
Based on this estimator, the marginal covariate-specific cumulative incidence function can be estimated by
\[
\hat{F}_{n,l}(t ; \boldsymbol{z}_{0})=\int_{0}^{t} \exp \left\{-\sum_{l=1}^{k} \hat{\Lambda}_{n,l}(u-; \boldsymbol{z}_{0})\right\} d \hat{\Lambda}_{n,l}(u ; \boldsymbol{z}_{0}),
\]
where $\hat{\Lambda}_{n,l}(t;\boldsymbol{z}_{0})=\hat{\Lambda}_{n,l}(t) \exp (\hat{\boldsymbol{\beta}}_{n,l}^{T} \boldsymbol{z}_{0})$.

\subsection{Asymptotic Properties}
\label{ss:asymp}

Here, we state the main theorems for the asymptotic properties of the proposed estimators $\hat{\boldsymbol{\beta}}_{n,l}$, $\hat{\Lambda}_{n,l}(t)$ and $\hat{F}_{n,l}(t ; \boldsymbol{z}_{0})$. The detailed proofs of these theorems are provided in Web Appendix B. For simplicity, we will omit the subindex $i$, indicating a specific cluster, from expectations. By the i.i.d. assumption across clusters, the expectations correspond to expectations of (functions of) random variables from an arbitrary cluster. These expectations Before stating the regularity conditions, we define the negative of the true pseudo-Hessian matrix as
\[
\boldsymbol{H}_{l}\left(\boldsymbol{\beta}\right)=E\left\{\frac{1}{M} \sum_{j=1}^{M} \int_{0}^{\tau} \boldsymbol{V} \left(t, \boldsymbol{\beta} \right) d \tilde{N}_{j l}\left(t ; \boldsymbol{\gamma}_{0}\right)\right\},
\]
where
\[
\boldsymbol{V}(t, \boldsymbol{\beta})=
\frac{E\{\frac{1}{M} \sum_{j=1}^{M} Y_{j}(t) \exp (\boldsymbol{\beta}^{T} \boldsymbol{Z}_{j}) \boldsymbol{Z}_{j}^{\otimes 2}\}}{E\{\frac{1}{M} \sum_{j=1}^{M} Y_{j}(t) \exp (\boldsymbol{\beta}^{T} \boldsymbol{Z}_{j})\}}-\left[\frac{E\{\frac{1}{M} \sum_{j=1}^{M} Y_{j}(t) \exp (\boldsymbol{\beta}^{T} \boldsymbol{Z}_{j}) \boldsymbol{Z}_{j}\}}{E\{\frac{1}{M} \sum_{j=1}^{M} Y_{j}(t) \exp (\boldsymbol{\beta}^{T} \boldsymbol{Z}_{j})\}}\right]^{\otimes 2}.
\]
The following regularity conditions are assumed throughout the remainder of this paper.
\begin{enumerate}
\item[C1.] $\Lambda_{0, l}(t)$ is a non-decreasing continuous function with $\Lambda_{0,l}(\tau)<\infty$, for $l=1, \ldots, k$ and $E\{Y(\tau) \mid \boldsymbol{Z}, M\}>0$ almost surely.
\item[C2.] The true regression coefficients $\boldsymbol{\beta}_{0,l} \in \mathcal{B}_{l} \subset \mathbb{R}^{p_{l}}$, where $\mathcal{B}_{l}$ is bounded and convex set for $l=1, \ldots, k$ and $\boldsymbol{\beta}_{0, l}$ is in interior of $\mathcal{B}_{l}$.
\item[C3.] The inverse of the link function $g$ for the marginal probability model of the cause of failure $\pi_{l}(\boldsymbol{W}_{i j}, \boldsymbol{\gamma}_{0})$, $l=1, \ldots, k$, has continuous derivative $\dot{g}$ with respect to $\boldsymbol{\gamma}_{0}$ on compact sets. The parameter space $\Gamma$ of $\boldsymbol{\gamma}_{0}$ is a bounded subset of $\mathbb{R}^{{p}_{\gamma}}$.
\item[C4.] The estimating function for the model of the cause of failure is Lipschitz continuous in  $\boldsymbol{\gamma}$, and the estimator $\hat{\boldsymbol{\gamma}}_n$ is strongly consistent and asymptotically linear, i.e. $\sqrt{n}(\hat{\boldsymbol{\gamma}}_n-\boldsymbol{\gamma}_{0})=\frac{1}{\sqrt{n}} \sum_{i=1}^{n} \frac{1}{M_{i}} \sum_{j=1}^{M_{i}}\boldsymbol{\omega}_{i j}+o_{p}(1)$, where
$\boldsymbol{\omega}_{i j}$ is the influence function of $j$th subject in $i$th cluster, satisfying $E(\boldsymbol{\omega}_{i j})=0$ and $E\left\|\boldsymbol{\omega}_{i j}\right\|^{2}<\infty$.
\item[C5.] The covariates of interest $\boldsymbol{Z}$, the auxiliary covariates $\boldsymbol{A}$, and the cluster size $M$ are bounded, in the sense that there exist constants $K \in \mathbb{R}_{+}$ and $m_{0} \in \mathbb{N}_{+}$ such that $P(\|{\boldsymbol{Z}}\| \vee \|{\boldsymbol{A}}\|\leq K)=1$ and $P(M \leq m_{0})=1$.
\item[C6.] The true pseudo-Hessian matrix $-\boldsymbol{H}_{l}(\boldsymbol{\beta})$ is negative definite on $\mathcal{B}_l$ for all $l=1, \ldots, k$.
\item[C7.] $\boldsymbol{Z}_{ij}$, $Y_{ij}(t)$, and $N_{ijl}(t)$ are identically distributed conditionally on cluster size $M_i$, in the sense that $E(\boldsymbol{Z}_{ij}|M_i)=E(\boldsymbol{Z}_{i1}|M_i)$, $E\{Y_{ij}(t)|M_i\}=E(Y_{i1}(t)|M_i)$, and $E\{N_{ijl}(t)|M_i\}=E(N_{i1l}(t)|M_i)$, for all $i=1,\ldots,n$, $j=1,\ldots,M_i$, and $l=1,\ldots,k$.
\end{enumerate}
Regularity conditions C3 and C4 are satisfied when the marginal model for $\pi_{l}(\boldsymbol{W}_{i j}, \boldsymbol{\gamma}_{0})$ is correctly specified with a standard link function, such as the logit link, and parameters estimated through generalized estimating equations under a working independence assumption.
The assumptions on the parametric models for $\pi_l(\boldsymbol{W}_{i j},\boldsymbol{\gamma}_{0})$ can be evaluated using the cumulative residual processes
\[
E\left[\frac{1}{M}\sum_{j=1}^{M} R_{j}\{N_{j l}(t)-\pi_l(\boldsymbol{W}_{ j},\boldsymbol{\gamma}_{0})N_{j}(t)\}\right], \ \ \ \ l=1, \ldots, k-1, \ \ t \in [0,\tau],
\]
which can be estimated by
\[
\frac{1}{n} \sum_{i=1}^{n} \frac{1}{M_{i}}\sum_{j=1}^{M_{i}} R_{i j}\{N_{i j l}(t)-\pi_l(\boldsymbol{W}_{i j},\hat{\boldsymbol{\gamma}}_{n})N_{ij}(t)\}.
\]
If the model is correctly specified, the cumulative residual process is equal to 0 for $t \in [0,\tau]$. A formal goodness of fit test can be conducted using the simulation-based approach by \citet{Pan05}. A graphical evaluation of goodness of fit can also be performed by plotting the observed residual process and the 95\% simultaneous confidence band around the line $f(t)=0$, $t \in [0,\tau]$ \citep{Bakoyannis19, Bakoyannis20}.

Theorem \ref{thm1} states the consistency of the proposed estimators $\hat{\boldsymbol{\beta}}_{n,l}$ and $\hat{\Lambda}_{n,l}(t)$. The proof of Theorem \ref{thm1} is given in Web Appendix B.1.
\begin{thm}
\label{thm1}
Under the assumptions in Section~\ref{ss:notation} and regularity conditions C1 - C7,
\[
\sum_{l=1}^{k}\left\{\|\hat{\boldsymbol{\beta}}_{n,l}-\boldsymbol{\beta}_{0,l}\|+\| \hat{\Lambda}_{n,l}(t)-\Lambda_{0,l}(t) \|_{\infty}\right\} \rightarrow_{as^{*}} 0,
\]
as $n\rightarrow\infty$, where $\|f(t)\|_{\infty}=\sup_{ t\in [0,\tau]}|f(t)|$.
\end{thm}
A corollary of Theorem \ref{thm1} is the strong uniform consistency of $\hat{F}_{n,l}(t ; \boldsymbol{z}_{0})$, $l=1,\ldots,k$, that is $\sum_{l=1}^{k}\|\hat{F}_{n,l}(t ; \boldsymbol{z}_{0})-F_{0,l}(t ; \boldsymbol{z}_{0})\|_{\infty} \rightarrow_{a s^{*}} 0$.

Theorem \ref{thm2} provides the asymptotic distribution for the finite-dimensional parameter $\hat{\boldsymbol{\beta}}_{n, l}$, which provides the basis for statistical inference about the regression coefficients $\boldsymbol{\beta}_{0,l}$, for $l=1, \ldots, k$. The proof of Theorem 2 is given in Web Appendix B.2. Before providing the theorem, we define the following quantities
\[
\boldsymbol{\psi}_{i j l}={\boldsymbol{H}}_{l}^{-1}(\boldsymbol{\beta}_{0,l})\int_0^{\tau}\{{\boldsymbol{Z}}_{i j}-\boldsymbol{E}(t,\boldsymbol{\beta}_{0,l})\}d\tilde{M}_{i j l}(t;\boldsymbol{\beta}_{0,l},\boldsymbol{\gamma}_0),
\]
where
\[
\boldsymbol{E}(t, \boldsymbol{\beta}_{0,l})=\frac{E\{\frac{1}{M}\sum_{j=1}^{M} Y_{j}(t) \exp (\boldsymbol{\beta}_{0,l}^{T} \boldsymbol{Z}_{j}) \boldsymbol{Z}_{\boldsymbol{j}}\}}
{E\{\frac{1}{M}\sum_{j=1}^{M} Y_{j}(t) \exp (\boldsymbol{\beta}_{0,l}^{T} \boldsymbol{Z}_{j})\}},
\]
and $\tilde{M}_{i j l}(t ; \boldsymbol{\beta}_{0, l}, \boldsymbol{\gamma}_{0})=\tilde{N}_{i j l}(t ; \boldsymbol{\gamma}_{0})-\int_{0}^{t} Y_{i j}(u) \exp (\boldsymbol{\beta}_{0, l}^{T} \boldsymbol{Z}_{i j}) d \Lambda_{0, l}(u)$,
where
\[
\Lambda_{0, l}(t)=\int_{0}^{t} \frac{E\{\frac{1}{M} \sum_{j=1}^{M} d \tilde{N}_{j l}(u ; \boldsymbol{\gamma}_{0})\}}{E\{\frac{1}{M} \sum_{j=1}^{M} Y_{j}(u) \exp (\boldsymbol{\beta}_{0,l}^{T} \boldsymbol{Z}_{j})\}}.
\]
Finally, we define the non-random quantity
\[
\boldsymbol{R}_{l}=\boldsymbol{H}_{l}^{-1}(\boldsymbol{\beta}_{0, l}) E\left[\frac{1}{M} \sum_{j=1}^{M} (1-R_{j})\int_{0}^{\tau}\{\boldsymbol{Z}_{j}-\boldsymbol{E}(t, \boldsymbol{\beta}_{0, l})\} d N_{j}(t)\dot{\pi}_{l}(\boldsymbol{W}_{j}, \boldsymbol{\gamma}_{0})^{T}\right],
\]
where $\dot{\pi}_{l}(\boldsymbol{W}_{j},\boldsymbol{\gamma}_{0}) = \partial\{\pi_{l}({\boldsymbol{W}_{j}},\boldsymbol{\gamma})\}(\partial\boldsymbol{\gamma})^{-1}|_{\boldsymbol{\gamma}=\boldsymbol{\gamma}_0}$.
\begin{thm}
\label{thm2}
Under the assumptions in Section~\ref{ss:notation} and regularity conditions C1 - C7, for $l=1, \ldots, k$,
$
\sqrt{n}(\hat{\boldsymbol{\beta}}_{n, l}-\boldsymbol{\beta}_{0, l})=n^{-1/2} \sum_{i=1}^{n}\left\{M_{i}^{-1} \sum_{j=1}^{M_{i}}\left(\boldsymbol{\psi}_{i j l}+\boldsymbol{R}_{l}\boldsymbol{\omega}_{i j}\right)\right\}+o_{p}(1)$.
\end{thm}
By Theorem 2, $\sqrt{n}(\hat{\boldsymbol{\beta}}_{n, l}-\boldsymbol{\beta}_{0, l}) \rightarrow_{d} N(\boldsymbol{0}, \boldsymbol{\Sigma}_{l})$, where
$
\boldsymbol{\Sigma}_{l}=E\{M^{-1} \sum_{j=1}^{M}(\boldsymbol{\psi}_{j l}+\boldsymbol{R}_{l}\boldsymbol{\omega}_{j})\}^{\otimes 2}$. The covariance matrix $\boldsymbol{\Sigma}_{l}$ can be consistently estimated using the empirical versions of the influence functions by
\[
\hat{\boldsymbol{\Sigma}}_{l}=\frac{1}{n} \sum_{i=1}^{n} \left\{\frac{1}{M_{i}} \sum_{j=1}^{M_{i}}\left(\hat{\boldsymbol{\psi}}_{i j l}+\hat{\boldsymbol{R}}_{l}\hat{\boldsymbol{\omega}}_{i j}\right)\right\}^{\otimes 2}.
\]
The empirical versions of the influence functions can be obtained by replacing expectations with sample averages over clusters and unknown parameters with their consistent estimates. Explicit formulas for the empirical versions of the influence functions are provided in Web Appendix B.5.

Theorems \ref{thm3} and \ref{thm4} provide the weak convergence of $\hat{\Lambda}_{n,l}(t)$ and $\hat{F}_{n,l}(t ; \boldsymbol{z}_{0})$, respectively. 
Before providing these theorems, we define some useful quantities that appear in the influence functions of the estimators of the infinite-dimensional parameters. For $l=1,\ldots,k$ and $t\in[0,\tau]$, define
\[
\phi_{ijl}(t)=\int_0^t\frac{d\tilde{M}_{ijl}(s;\boldsymbol{\beta}_{0,j},\boldsymbol{\gamma}_0)}{E\{\frac{1}{M} \sum_{j=1}^{M} Y_{j}(s) \exp (\boldsymbol{\beta}_{0,l}^{T} \boldsymbol{Z}_{j})\}}-(\boldsymbol{\psi}_{ijl}+{\boldsymbol{R}}_l\boldsymbol{\omega}_{ij})^T\int_0^t \boldsymbol{E}(s,\boldsymbol{\beta}_{0,l})d\Lambda_{0,l}(s),
\]
and the non-random function
\[
{\boldsymbol{R}}_j^{\star}(t)=E\left[\frac{1}{M} \sum_{j=1}^{M} (1-R_{j})\dot{\pi}_j({\boldsymbol{W}_{j}},\boldsymbol{\gamma}_0)\int_0^t\frac{dN_{j}(s)}{E\{\frac{1}{M} \sum_{j=1}^{M} Y_{j}(s) \exp (\boldsymbol{\beta}_{0,l}^{T} \boldsymbol{Z}_{j})\}}\right]^T.
\]
In addition, we define the influence function
\begin{eqnarray*}
& &\phi_{ijl}^F(t;{\boldsymbol{z}}_0)=\int_0^t\exp\left\{-\sum_{l=1}^k\Lambda_{0,l}(s-;{\boldsymbol{z}}_0)\right\}d\phi^{\Lambda}_{ijl}(s;{\boldsymbol{z}}_0)\\
& &\hspace{25mm}-\int_0^t\left\{\sum_{l=1}^k\phi^{\Lambda}_{ijl}(s-;{\boldsymbol{z}}_0)\right\}\exp\left\{-\sum_{l=1}^k\Lambda_{0,l}(s-;{\boldsymbol{z}}_0)\right\}d\Lambda_{0,l}(s;{\boldsymbol{z}}_0),
\end{eqnarray*}
where
$\phi_{ijl}^{\Lambda}(t;{\boldsymbol{z}}_0)=\{{\boldsymbol{z}}_0^T(\boldsymbol{\psi}_{ijl}+{\boldsymbol{R}}_l\boldsymbol{\omega}_{ij})\Lambda_{0,l}(t)+\phi_{ijl}(t)+{\boldsymbol{R}}_l^{\star}(t)\boldsymbol{\omega}_{ij}\}\exp(\boldsymbol{\beta}_{0,l}^T{\boldsymbol{z}}_0)$. Finally, $D[0,\tau]$ denotes the space of right-continuous functions with left-hand limits defined on $[0, \tau]$, and $\{\xi_{i}\}_{i=1}^{n}$ are standard normal variables
independent of the data. The proofs of the following theorems are given in Web Appendices B.3 and B.4.

\begin{thm}
\label{thm3}
Under the assumptions in Section~\ref{ss:notation} and regularity conditions C1 - C7, for $l=1, \ldots, k$, and $t\in[0,\tau]$,
\[
\sqrt{n}\left\{\hat{\Lambda}_{n, l}(t)-\Lambda_{0, l}(t)\right\}=\frac{1}{\sqrt{n}} \sum_{i=1}^{n}\left[ \frac{1}{M_{i}} \sum_{j=1}^{M_{i}} \left\{\phi_{i j l}(t)+\boldsymbol{R}_{l}^{*}(t) \boldsymbol{\omega}_{i j}\right\}\right]+o_{p}(1),
\]
with the influence functions belonging to a Donsker class, and conditional on the observed data, $\hat{W}_{n, l}(\cdot)=n^{-1/2} \sum_{i=1}^{n}[ M_{i}^{-1} \sum_{j=1}^{M_{i}}\{\hat{\phi}_{i j l}(\cdot)+\hat{\boldsymbol{R}}_{l}^{*}(\cdot) \hat{\boldsymbol{\omega}}_{i j}\}]\xi_{i}$ converges weakly to the same limiting process as $W_{n, l}(\cdot)=\sqrt{n}\{\hat{\Lambda}_{n, l}(\cdot)-\Lambda_{0, l}(\cdot)\}$.
\end{thm}
By Theorem 3, $\sqrt{n}\{\hat{\Lambda}_{n, l}(\cdot)-\Lambda_{0, l}(\cdot)\}\leadsto \mathbb{G}_{\Lambda_l}$ in $D[0,\tau]$, where $\mathbb{G}_{\Lambda_l}$ is a tight mean zero Gaussian process with covariance function $E[ M^{-1} \sum_{j=1}^{M} \{\phi_{j l}(t)+\boldsymbol{R}_{l}^{*}(t) \boldsymbol{\omega}_{j}\}][ M^{-1} \sum_{j=1}^{M} \{\phi_{j l}(s)+\boldsymbol{R}_{l}^{*}(s) \boldsymbol{\omega}_{j}\}]$, $t,s \in [0, \tau]$. A consistent estimator of the covariance function is
\[
\frac{1}{n} \sum_{i=1}^{n}\left[\frac{1}{M_{i}} \sum_{j=1}^{M_{i}}\left\{\hat{\phi}_{i j l}(t)+\hat{\boldsymbol{R}}_{l}^{*}(t) \hat{\boldsymbol{\omega}}_{i j}\right\}\right]\left[\frac{1}{M_{i}} \sum_{j=1}^{M_{i}}\left\{\hat{\phi}_{i j l}(s)+\hat{\boldsymbol{R}}_{l}^{*}(s) \hat{\boldsymbol{\omega}}_{i j}\right\}\right], \quad\ t,s \in [0, \tau].
\]
Explicit formulas for the empirical versions of the influence functions are provided in Web Appendix B.5. Calculation of confidence intervals and bands can be performed using an appropriate continuously differentiable transformation to avoid negative limits \citep{Lin94}. A standard choice is the transformation $g(x)=\log (x)$. According to the functional delta method, $\sqrt{n} q_{l}^{\Lambda}(t)[g\{\hat{\Lambda}_{n, l}(t)\}-g\{\Lambda_{0, l}(t)\}]$ is asymptotically equivalent to $B_{n, l}(t)=q_{l}^{\Lambda}(t) \dot{g}\{\hat{\Lambda}_{n, l}(t)\} W_{n, l}(t)$. Also, by Theorem \ref{thm3}, $B_{n, l}(t)$ is asymptotically equivalent to $\hat{B}_{n, l}(t)=q_{l}^{\Lambda}(t) \dot{g}\{\hat{\Lambda}_{n, l}(t)\} \hat{W}_{n, l}(t)$, where $q_{l}^{\Lambda}(t)$ is a weight function, with $t \in[t_{1}, t_{2}]$, $0 \leq t_{1} \leq t_{2} < \tau$. The choice $q_{l}^{\Lambda}(t)=\hat{\Lambda}_{n, l}(t)\{\hat{\sigma}_{\Lambda_{l}}(t)\}^{-1}$, where $\hat{\sigma}_{\Lambda_{l}}(t)$ is the square root of the estimated variance of $\hat{\Lambda}_{n, l}(t)$, gives the equal precision band \citep{Nair84}; the choice $q_{l}^{\Lambda}(t)=\hat{\Lambda}_{n, l}(t)\{1+\hat{\sigma}^{2}_{\Lambda_{l}}(t)\}^{-1}$, provides a Hall-Wellner-type band \citep{Hall80}. Now, a $1-\alpha$ confidence band for $\Lambda_{0, l}(t)$ can be computed as
\[
g^{-1}\left[g\left\{\hat{\Lambda}_{n, l}(t)\right\} \pm \frac{c_{\alpha}}{\sqrt{n} q_{l}^{\Lambda}(t)}\right], \quad\ t \in[t_{1}, t_{2}],
\]
where $c_{\alpha}$ is the $1-\alpha$ quantile of the distribution of $\sup _{t \in[t_{1}, t_{2}]}|\hat{B}_{n,l}(t)|$. This can be estimated using a large number of simulation realizations from the process $\hat{B}_{n,l}(\cdot)$, generated by repeatedly simulating sets of standard normal variables $\{\xi_i\}_{i=1}^n$ \citep{Spiekerman98}. Since confidence bands tend to be unstable at earlier and later time points, where there are fewer observed events, we suggest the restriction of the confidence band domain $[t_{1}, t_{2}]$ to the 10th and 90th percentile of the event times.
\begin{thm}
\label{thm4}
Under the assumptions in Section~\ref{ss:notation} and regularity conditions C1 - C7, for $l=1, \ldots, k$, and $t\in[0,\tau]$,
\[
\sqrt{n}\left\{\hat{F}_{n,l}(t ; \boldsymbol{z}_{0})-F_{0,l}(t ; \boldsymbol{z}_{0})\right\}=\frac{1}{\sqrt{n}}\sum_{i=1}^{n} \left\{\frac{1}{M_{i}} \sum_{j=1}^{M_{i}}\phi_{i j l}^{F}(t ; \boldsymbol{z}_{0})\right\}+o_{p}(1),
\]
with the influence functions belonging to a Donsker class, and conditionally on the observed data, $\hat{W}_{n, l}^{F}(\cdot ; \boldsymbol{z}_{0})=n^{-1/2} \sum_{i=1}^{n} \{ M_{i}^{-1} \sum_{j=1}^{M_{i}} \hat{\phi}_{i j l}^{F}(\cdot ; \boldsymbol{z}_{0})\} \xi_{i}$ converges weakly to the same limiting process as $W_{n, l}^{F}(\cdot ; \boldsymbol{z}_{0})=\sqrt{n}\{\hat{F}_{n,l}(\cdot ; \boldsymbol{z}_{0})-F_{0,l}(\cdot ; \boldsymbol{z}_{0})\}$.
\end{thm}

By Theorem 4, $\sqrt{n}\{\hat{F}_{n,l}(\cdot ; \boldsymbol{z}_{0})-F_{0,l}(\cdot ; \boldsymbol{z}_{0})\}\leadsto \mathbb{G}_{F_l}$ in $D[0,\tau]$, where $\mathbb{G}_{F_l}$ is a tight mean zero Gaussian process with covariance function $E\{ M^{-1} \sum_{j=1}^{M} \phi_{j l}^{F}(t ; \boldsymbol{z}_{0})\}\{ M^{-1} \sum_{j=1}^{M} \phi_{j l}^{F}(s ; \boldsymbol{z}_{0})\}$.
A consistent estimator for the covariance function is
\[
\frac{1}{n} \sum_{i=1}^{n}\left\{ \frac{1}{M_{i}} \sum_{j=1}^{M_{i}} \hat{\phi}_{i j l}^{F}(t ; \boldsymbol{z}_{0})\right\}\left\{ \frac{1}{M_{i}} \sum_{j=1}^{M_{i}} \hat{\phi}_{i j l}^{F}(s; \boldsymbol{z}_{0})\right\}, \quad\ t, s \in[0, \tau].
\]
Explicit formulas for the empirical versions of the influence functions are provided in Web Appendix B.5. Similarly to the case of the cumulative baseline hazards $\Lambda_{0,l}$, a $1-\alpha$ confidence band for $F_{0,l}(\cdot ; \boldsymbol{z}_{0})$ can be constructed as
\[
g^{-1} \left[g\left\{\hat{F}_{0,l}(t ; \boldsymbol{z}_{0})\right\} \pm \frac{c_{\alpha}}{\sqrt{n} q_{l}^{F}\left(t ; \boldsymbol{z}_{0}\right)}\right], \quad\ t \in [t_{1}, t_{2}],
\]
where $c_{\alpha}$ is the $1-\alpha$ quantile of the distribution of $\sup _{t \in[t_{1}, t_{2}]} |\hat{B}_{n, l}^{F}(t ; \boldsymbol{z}_{0})|$, with $\hat{B}_{n, l}^{F}(t ; \boldsymbol{z}_{0})=q_{l}^{F}(t ; \boldsymbol{z}_{0}) \dot{g}\{\hat{F}_{n, l}(t ; \boldsymbol{z}_{0})\} \hat{W}_{n, l}^{F}(t ; \boldsymbol{z}_{0})$. A standard transformation to ensure that the limits of the bands for the cumulative incidence functions reside in $[0,1]$ is $g(x)=\log \{-\log (x)\}$ \citep{Cheng98}. The weight function choice $q_{l}^{F}(t ; \boldsymbol{z}_{0})=\hat{F}_{n,l}(t ; \boldsymbol{z}_{0}) \log \{\hat{F}_{n,l}(t ; \boldsymbol{z}_{0})\}\{\hat{\sigma}_{F_{l}}(t ; \boldsymbol{z}_{0})\}^{-1}$, with $\hat{\sigma}_{F_{l}}\left(t ; \boldsymbol{z}_{0}\right)$ being the square root of the estimated variance of $\hat{F}_{0,l}(t ; \boldsymbol{z}_{0})$, provides an equal-precision-type band \citep{Nair84}; the choice $q_{l}^{F}(t ; \boldsymbol{z}_{0})=\hat{F}_{n,l}(t ; \boldsymbol{z}_{0}) \log \{\hat{F}_{n,l}(t ; \boldsymbol{z}_{0})\}\{1+\hat{\sigma}^{2}_{F_{l}} (t ; \boldsymbol{z}_{0})\}^{-1}$, provides a Hall-Wellner-type band \citep{Hall80}.

\section{Simulation Studies}
\label{s:sim}
To evaluate the finite-sample performance of the proposed estimators, and compare them with the estimators by \citet{Bakoyannis20} which ignore the within-cluster dependence, we conducted a series of simulation experiments. We considered simulation settings similar to those used in \citet{Bakoyannis20}, with two competing risks, two covariates $\boldsymbol{Z}=(Z_{1}, Z_{2})^{T}$, where $Z_1 \sim N(0,2^2)$ and $Z_2\sim \textrm{Bernoulli}(0.5)$, and with the observation time interval being $[0,2]$. The right censoring times were independently generated from the $\textrm{Exp}(0.4)$ distribution. For each cause, the failure times were generated from Cox proportional hazards shared frailty models with a positive stable frailty \citep{ Hougaard86, Cong07, Liu11} to introduce within-cluster dependence. These models had the form
\[
\lambda_{l}(t \mid Z_{ij,l},w_{i,l})=\lambda_{0,l}(t)w_{i,l}\exp(\beta_{0,l} Z_{ij,l}), \quad\ l=1,2,
\]
where $w_{i,l}$ followed a positive stable distribution with parameter $\alpha=0.5$, which induced a moderate within-cluster dependence. 
This simulation setup led to the marginal cause-specific hazard functions 
\[
\lambda_{l}(t \mid {Z}_{ij,l})=\alpha \lambda_{0,l}(t)\Lambda_{0,l}(t)^{\alpha-1}\exp(\alpha{\beta}_{0,l}{Z}_{ij,l}), \quad\ l=1,2,
\]
which were still proportional, owing to the positive stable frailty, with true parameters ${\beta}^{\prime}_{0,l}=\alpha{\beta}_{0,l}$, and $\Lambda_{0,l}^{\prime}(t)={\Lambda_{0,l}(t)}^{\alpha}$.

In this simulation study we considered two scenarios. In both scenarios, the event time for the cause $1$ was generated assuming $\lambda_{0,1}(t)=1$ and $\beta_{0,1}=-0.5$. In Scenario 1, the event time for cause $2$ was simulated from a Gompertz distribution with baseline hazard $\lambda_{0,1}(t)=\exp(-0.5+0.2t)$, and $\beta_{0,2}=-0.5$. In Scenario 2, the event time for cause $2$ was simulated from a Weibull distribution with baseline hazard $\lambda_{0,1}(t)=\{2\sqrt{2t}\}^{-1}$ and $\beta_{0,2}=-0.5$. The simulation setup under Scenario 1 led to on average in $13.5\%$ right-censored observations, $50.4\%$ failures from cause 1, and $36.1\%$ failures from cause 2. The corresponding figures for Scenario 2 were $12.7\%$, $41.8\%$, and $45.5\%$. The implied model for $\pi_1(\boldsymbol{W}_{i j},\boldsymbol{\gamma}_0)$ had approximately linear time effect with the form $\textrm{logit}\{\pi_1(\boldsymbol{W}_{i j},\boldsymbol{\gamma}_0)\}\approx\gamma_{0}+\gamma_{1}T+\gamma_{2}Z_{1}+\gamma_{3}Z_{2}$ under Scenario 1, where $\boldsymbol{\gamma}_0\approx(0.25,-0.15,-0.25,0.25)^T$, and had the form $\textrm{logit}\{\pi_1(\boldsymbol{W}_{i j},\boldsymbol{\gamma}_0)\}=\gamma_{0}+\gamma_{1}\log(T)+\gamma_{2}Z_{1}+\gamma_{3}Z_{2}$ under Scenario 2, where $\boldsymbol{\gamma}_0=(5\log(2)/4,0.25,-0.25,0.25)^T$. The missingness indicators $R_{ij}$ were generated under the logistic model
\[
\textrm{logit}\left\{P(R_{ij}=1\mid\Delta_{ij}=1,\boldsymbol{W}_{ij})\right\}=(1,\boldsymbol{W}^{T}_{ij})\boldsymbol{\theta},
\]
where $\boldsymbol{W}^{T}=(T,Z_{1},Z_{2})$. We considered the parameter values $\boldsymbol{\theta}=(0.7,1,-1,1)^{T}$, $(-0.2,1,-1,1)^{T}$, and $(-0.8,1,-1,1)^{T}$, which resulted in $24.5\%$, $35.1\%$, and $42.8\%$ missing causes of failure in Scenario 1, and $25.5\%$, $36.3\%$, and $44.1\%$ missingness in Scenario 2.

In this simulation study we considered $n=50$, $100$, or $200$ which correspond to situations with small to moderate number of clusters. To introduce informative cluster size, the cluster sizes $M_i$ were generated from a mixture of discrete uniform distributions depending on the frailty, with $M_i \sim \textrm{Unif}(20,30)$ if $w_{i,1}<\textrm{median}(w_1)$ and $w_{i,2}<\textrm{median}(w_2)$,
$M_i \sim \textrm{Unif}(50,60)$ if $w_{i,1}\ge \textrm{median}(w_1)$ and $w_{i,2}\ge\textrm{median}(w_2)$, and $M_i \sim \textrm{Unif}(30,50)$, otherwise.

For each simulation setting, we simulated 1000 datasets, and analyzed each dataset using the proposed method and the method by \citet{Bakoyannis20}. All analysis assumed the parametric model $\textrm{logit}\{\pi_1(\boldsymbol{W}_{i j},\boldsymbol{\gamma}_0)\}=\gamma_{0}+\gamma_{1}T+\gamma_{2}Z_{1}+\gamma_{3}Z_{2}$, which was approximately correctly specified under Scenario 1, and misspecified under Scenario 2. The standard errors were estimated using the closed-form formulas provided in Section~\ref{s:method}. The $95\%$ confidence bands for $\Lambda_{0,1}(t)$ and $F_{0,1}(t)$ were computed based on 1000 simulation realizations standard normal variables $\left\{\xi_{i}\right\}_{i=1}^{n}$ as described in Section~\ref{s:method}. The limits of the time domain $[t_1,t_2]$ for the confidence bands were chosen to be $10\%$ and $90\%$ percentile of the observed failure times.

The simulation results for the regression coefficient $\beta_{1}$ under Scenario 1 are summarized in Table~\ref{t:one}. The proposed estimator was approximately unbiased and the average standard errors were close to the Monte Carlo standard deviations. This provides numerical evidence for the consistency of our estimator of the regression coefficient $\hat{\beta}_{n,1}$ and its associated standard error. The $95\%$ coverage probabilities were close to the nominal level in all cases. In contrast, the method by \citet{Bakoyannis20} provided biased estimates. The bias for $\beta_{1}$ was relatively small under Scenario 1, with an increasing trend as the number of clusters increased. The average standard errors were smaller than the Monte Carlo standard deviation, which implies that the standard errors were under-estimated. This resulted in poor coverage probabilities of the corresponding $95\%$ confidence intervals.

\begin{table}
\caption{Simulation results for the regression coefficient $\beta_{1}$ under Scenario 1 for the proposed approach and the approach by \citet{Bakoyannis20} (BZY20) which ignores the within-cluster dependence.}
\label{t:one}
\begin{center}
\begin{tabular}{llcccccccc}
\hline
& & \multicolumn{4}{c}{Proposed} & \multicolumn{4}{c}{BZY20}\\
\cmidrule(lr){3-6}\cmidrule(lr){7-10}
$n$ & $p_m(\%)$ & Bias & MCSD & ASE & CP & Bias & MCSD & ASE & CP \\
\hline
50	&25	&-0.006	&0.033	&0.032	&0.937	&0.003	&0.033	&0.021	&0.782\\
	&35	&-0.006	&0.034	&0.033	&0.938	&0.003	&0.035	&0.023	&0.793\\
	&43	&-0.006	&0.036	&0.034	&0.939	&0.004	&0.035	&0.025	&0.827\\
100	&25	&-0.002	&0.022	&0.022	&0.949	&0.007	&0.022	&0.015	&0.777\\
	&35	&-0.002	&0.023	&0.023	&0.941	&0.007	&0.023	&0.016	&0.803\\
	&43	&-0.002	&0.024	&0.024	&0.948	&0.007	&0.024	&0.018	&0.822\\
200	&25	&-0.001	&0.016	&0.016	&0.954	&0.008	&0.016	&0.010	&0.735\\
	&35	&-0.001	&0.017	&0.017	&0.953	&0.008	&0.017	&0.011	&0.762\\
	&43	&-0.001	&0.017	&0.017	&0.953	&0.009	&0.017	&0.013	&0.785\\
\hline\\
\end{tabular}
\end{center}
{\footnotesize Note: $n$: number of clusters with cluster size $M \in [30,60]$; $p_m$: percentage of missingness; MCSD: Monte Carlo standard deviation; ASE: average estimated standard error; CP: coverage probability of $95\%$ confidence interval}
\end{table}

The simulation results for the pointwise estimates of the infinite-dimensional parameters $\Lambda_{0,1}(t)$ and $F_{0,1}(t)$ under Scenario 1 are provided in Web Appendix C. Our proposed estimators had good performance with small bias, average standard errors close to the Monte Carlo standard deviation, and $95\%$ confidence interval coverage probabilities close to the nominal level. As expected, the method by \citet{Bakoyannis20}, which ignores the within-cluster dependence and the informative cluster size, provided estimators with large bias, severely under-estimated standard errors, and poor coverage probabilities of the $95\%$ confidence intervals.

Table~\ref{t:three} presents the coverage probabilities of $95\%$ simultaneous confidence bands for the infinite-dimensional parameters $\Lambda_{0,1}(t)$ and $F_{0,1}(t)$ under Scenario 1. The proposed $95\%$ simultaneous confidence bands had coverage probabilities close to the nominal level, while the $95\%$ simultaneous confidence bands by \citet{Bakoyannis20} had a very poor coverage rate.

\begin{table}
\caption{Simulation results for the coverage probabilities of $95\%$ simultaneous confidence bands for the infinite-dimensional parameters $\Lambda_{0,1}(t)$ and $F_{0,1}(t)$ under Scenario 1. Results from the proposed approach and the approach by \citet{Bakoyannis20} (BZY20) which ignores the within-cluster dependence.}
\label{t:three}
\begin{center}
\begin{tabular}{llcccccccc}
\hline
& & \multicolumn{4}{c}{$\Lambda_{0,1}(t)$} & \multicolumn{4}{c}{$F_{0,1}(t)$}\\
\cmidrule(lr){3-6}\cmidrule(lr){7-10}
$n$ & $p_m(\%)$ & \multicolumn{2}{c}{Proposed} & \multicolumn{2}{c}{BZY20} & \multicolumn{2}{c}{Proposed} & \multicolumn{2}{c}{BZY20} \\
\cmidrule(lr){3-4}\cmidrule(lr){5-6}\cmidrule(lr){7-8}\cmidrule(lr){9-10}			
& & EP & HW & EP & HW & EP & HW & EP & HW \\
\hline
50	&25	&0.900	&0.936	&0.077	&0.153	&0.906	&0.931	&0.120	&0.203\\
	&35	&0.912	&0.936	&0.105	&0.185	&0.904	&0.928	&0.146	&0.245\\
	&43	&0.914	&0.939	&0.130	&0.213	&0.911	&0.929	&0.167	&0.268\\
100	&25	&0.931	&0.945	&0.049	&0.092	&0.931	&0.952	&0.096	&0.150\\
	&35	&0.931	&0.947	&0.064	&0.114	&0.932	&0.951	&0.106	&0.173\\
	&43	&0.937	&0.948	&0.082	&0.132	&0.939	&0.953	&0.141	&0.210\\
200	&25	&0.942	&0.947	&0.016	&0.034	&0.938	&0.955	&0.073	&0.108\\
	&35	&0.940	&0.950	&0.025	&0.046	&0.945	&0.954	&0.085	&0.129\\
	&43	&0.942	&0.951	&0.041	&0.062	&0.945	&0.955	&0.105	&0.154\\
\hline\\
\end{tabular}
\end{center}
{\footnotesize Note: $n$: number of clusters with cluster size $M \in [30,60]$; $p_m$: percentage of missingness; EP: equal precision bands; HW: Hall-Wellner-type bands}
\end{table}

The simulation results under Scenario 2, where the model for $\pi_1(\boldsymbol{W}_{i j},\boldsymbol{\gamma})$ was misspecified,  are provided in Web Appendix C. The results for point estimates under this scenario were similar to those form Scenario 1 (Table~\ref{t:one}). This provides numerical evidence for the robustness of the proposed approach under some degree of model misspecification in $\pi_l(\boldsymbol{W}_{i j},\boldsymbol{\gamma})$. However, the confidence bands had lower coverage rate under Scenario 2. This illustrates the importance to evaluate the goodness of fit of the model assumption for $\pi_l(\boldsymbol{W}_{i j},\boldsymbol{\gamma})$ in practice, as described in Section~\ref{ss:asymp}.

\section{HIV Data Application}
\label{s:data}
The proposed method was applied to the electronic health record data from the EA-IeDEA study to identify factors affecting disengagement from HIV care and death while in care (i.e. prior to a disengagement). Disengagement from care and death while in care were the two competing risks of interest. Disengagement from care was defined by the clinical investigators of the study as being alive and without HIV care for two months. The covariates of interest were sex, age, CD4 count at ART initiation, and HIV status disclosure. The dataset included 24373 HIV-infected adult patients from 31 clinics who initiated ART on/after January 1, 2010. Among these patients, 8082 were still in care, 84 died while in care (reported death), and 16207 were lost to clinic for at least two months. Among those 16207 patients who were lost to clinic, 5107 (31.5\%) were intensively outreached in the community and their vital status was actively ascertained by outreach workers. Among them, 1867 (36.6\%) patients were found to be deceased within two months from the last clinic visit, which indicates a substantial death under-reporting issue. The remaining 11100 lost patients who were not outreached had a missing cause of failure. Descriptive characteristics of the patients included in this analysis are presented in Table~\ref{t:four}.

\begin{table}
\caption{Descriptive characteristics for the EA-IeDEA study sample included in the analysis}
\label{t:four}
\begin{center}
\begin{tabular}{lllll}
\hline
& Right censoring & \multicolumn{3}{c}{Cause of failure} \\
\cmidrule(lr){2-2}\cmidrule(lr){3-5}
Variable & In care & Disengagement & Death & Missing \\
 & ($N$=8082) & ($N$=3240) & ($N$=1951$^1$) & ($N$=11100) \\
\hline
&$n$ (\%) & $n$ (\%) & $n$ (\%) & $n$ (\%) \\
Gender&&&& \\
Female & 5334 (66.0) & 2002 (61.8) & 974 (49.9) & 7363 (66.3) \\
Male & 2748 (34.0) & 1238 (38.2) & 977 (50.1) & 3737 (33.7) \\
HIV status disclosed&&&& \\
Yes & 5116 (63.3) & 2022 (62.4) & 1417 (72.6) & 6917 (62.3) \\
No & 2966 (36.7) & 1218 (37.6) & 534 (27.4) & 4183 (37.7) \\
\\
& Median (IQR) & Median (IQR) & Median (IQR) & Median (IQR) \\
\\
Age$^2$ & 38.0 (31.5, 46.1) & 36.1 (29.4, 43.3) & 39.1 (32.3, 48.2) & 36.0 (29.3, 43.9) \\
CD4$^3$ & 196 (95, 297) & 173 (72, 281) & 66 (22, 168)   & 183 (82, 291) \\
\hline\\
\end{tabular}
\end{center}
{\footnotesize Note: $^1$: Included 84 reported deaths and 1867 unreported deaths confirmed through outreach; $^2$: Age at ART initiation in years; $^3$: CD4 count at ART initiation in cells/$\mu$l}
\end{table}

\begin{figure}
\begin{center}
 \includegraphics[width=0.6\textwidth]{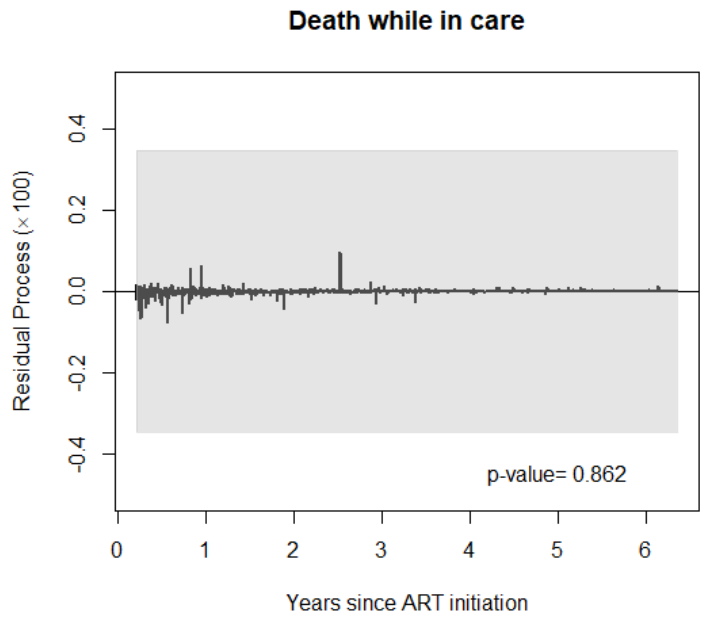}
\caption{Plot for the cumulative residual process of the parametric model $\pi_l(\boldsymbol{W}_{i j},\boldsymbol{\gamma}_0)$, $l=1,2$ for death while in care based on the HIV data along with the 95\% goodness-of-fit band (grey area) and the corresponding p-value.}
\label{f:one}
\end{center}
\end{figure}

We assumed a marginal logistic model for $\pi_l(\boldsymbol{W}_{i j},\boldsymbol{\gamma}_0)$, $l=1,2$ with time since ART initiation, sex, age, CD4 count at ART initiation, and HIV status disclosure as covariates. The goodness of fit evaluation for the parametric model $\pi_1(\boldsymbol{W}_{i j},\boldsymbol{\gamma}_0)$, is depicted in Figure~\ref{f:one}. The estimated residual process for death while in care fell within the 95\% simultaneous confidence band under the null hypothesis (p-value = 0.862). This indicates that there is no evidence for a violation of the parametric model assumption imposed for this dataset.

The data analysis results using the proposed method and the method by \citet{Bakoyannis20} are summarized in Table~\ref{t:five}. Younger patients with a higher CD4 count at ART initiation had a higher hazard of disengagement from HIV care, while males and patients with a lower CD4 count at ART initiation had a higher hazard of death while in care. In contrast, the method by \citet{Bakoyannis20} which ignores the within-cluster dependence and the informative cluster size, provided significant sex and HIV status disclosure effects on the hazard of disengagement from HIV care, and significant age and HIV status disclosure effects on the hazard of death while in care. The dubiously significant effects from the na\"ive analysis may be attributed to the under-estimation of standard errors, in addition to the bias due to the potential informative cluster size.

\begin{table}
\caption{Data analysis of the EA-IeDEA study sample. Results from the proposed approach and the approach by \citet{Bakoyannis20} (BZY20) which ignores the within-cluster dependence.}
\label{t:five}
\begin{center}
\begin{tabular}{lcccccc}
\hline
& \multicolumn{3}{c}{Proposed$^1$} & \multicolumn{3}{c}{BZY20$^2$}\\
\cmidrule(lr){2-4}\cmidrule(lr){5-7}
Covariates & $\exp(\hat{\beta}_n)$ & 95\% CI$^3$ & \textit{p}-value & $\exp(\hat{\beta}_n)$ & 95\% CI$^3$ & \textit{p}-value\\
\hline
Disengagement from HIV care & & & \\
\noalign{\smallskip}
Sex (male = 1, female = 0) & 0.94 & (0.80, 1.10) & 0.411 & 1.07 & (1.01, 1.12) & 0.016 \\
Age (10 years) & 0.79 & (0.74, 0.84) & $<$0.001 & 0.77 & (0.75, 0.79) & $<$0.001 \\
CD4 (100 cells/$\mu$l) & 1.06 & (1.00, 1.12) & 0.050 & 1.05 & (1.04, 1.06) & $<$0.001 \\
HIV status$^4$ & 0.96 & (0.84, 1.11) & 0.606 & 0.83 & (0.79, 0.87) & $<$0.001 \\
\hline
Death while in care & & & \\
\noalign{\smallskip}
Sex (male = 1, female = 0) & 1.40 & (1.20, 1.41) & $<$0.001 & 1.30 & (1.20, 1.41) & $<$0.001\\
Age (10 years) & 0.98 & (0.89, 1.09) & 0.758 & 1.07 & (1.03, 1.11) & 0.001\\
CD4 (100 cells/$\mu$l) & 0.63 & (0.57, 0.70) & $<$0.001 & 0.67 & (0.64, 0.71) & $<$0.001\\
HIV status$^4$ & 1.15 & (0.88, 1.50) & 0.310 & 1.27 & (1.16, 1.39) & $<$0.001 \\
\hline\\
\end{tabular}
\end{center}
{\footnotesize Note: $^1$: The standard errors were estimated with cluster bootstrap; $^2$: The standard errors were estimated with bootstrap at individual level; $^3$: 95\% confidence interval for the cause-specific hazard ratio $\exp(\beta_0)$; $^4$: HIV status disclosed with Yes = 1, No = 0}
\end{table}

\begin{figure}
 \includegraphics[width=0.95\textwidth]{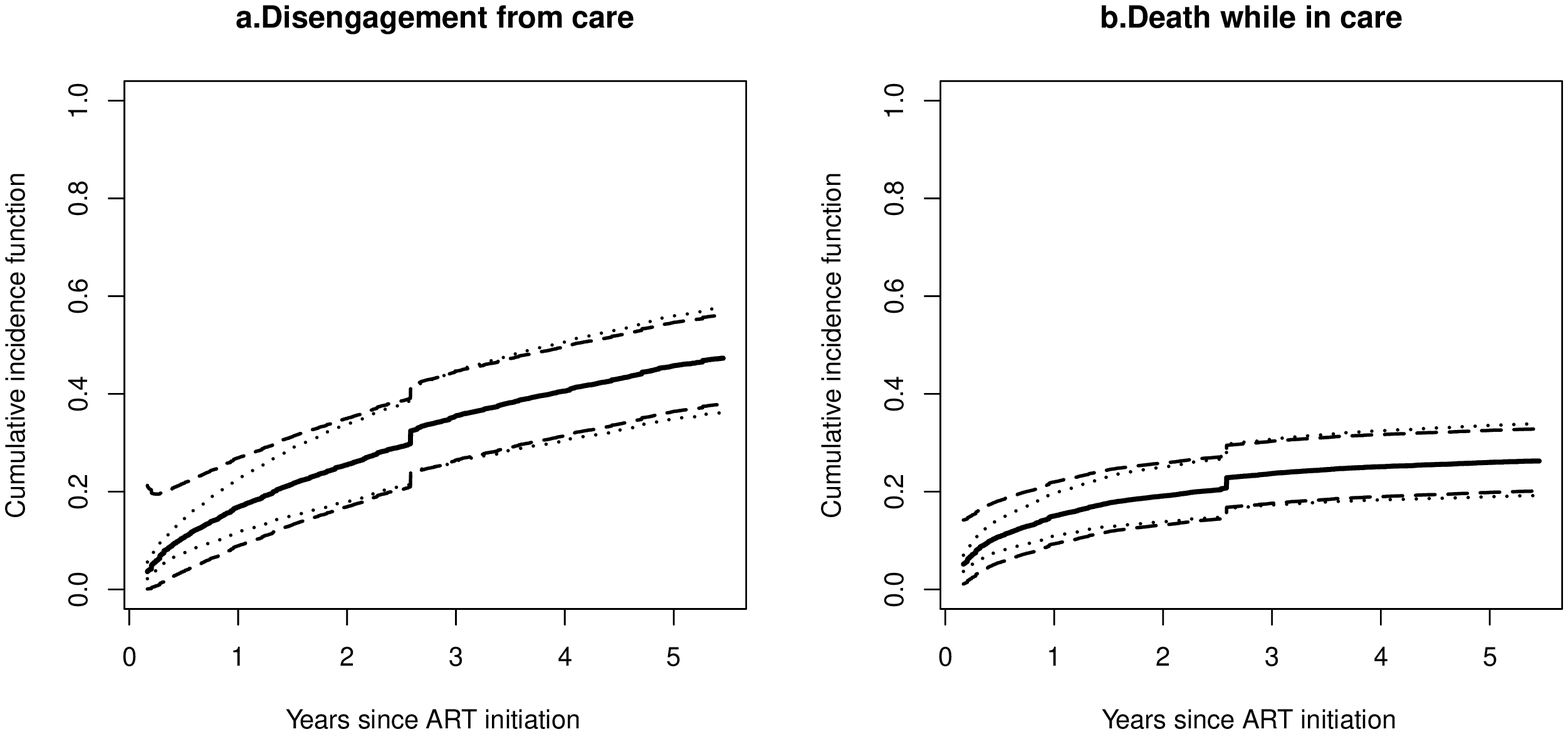}
\caption{Plot for predicted cumulative incidence functions (solid lines) of (a) disengagement from HIV care and (b) death while in care with the $95\%$ simultaneous confidence bands based on equal precision bands (dotted lines) and Hall-Wellner bands (dashed lines), for a 40-year-old man with CD4 count of 150 cells/$\mu$l at ART initiation and undisclosed HIV status}
\label{f:two}
\end{figure}

Figure~\ref{f:two} depicts the predicted cumulative incidence functions of (a) disengagement from care and (b) death while in care for a 40-year-old man with CD4 count of 150 cells/$\mu$l at ART initiation and undisclosed HIV status, along with the equal-precision-type and the Hall-Wellner-type bands.

\section{Discussion}
\label{s:discuss}

In this paper, we proposed a general framework for marginal semiparametric regression analysis of clustered competing risks data with missing cause of failure. Our approach utilizes the marginal proportional cause-specific hazards model, and uses a partial pseudolikelihood approach for estimation under a MAR assumption. We provide estimators for both regression coefficients and infinite-dimensional parameters, such as the marginal cumulative incidence function. The proposed method does not impose assumptions regarding the within-cluster dependence and also accounts for informative cluster size. The proposed estimators were shown to be strongly consistent and asymptotically normal. Closed-form variance estimators were provided and rigorous methodology for the calculation of simultaneous confidence bands for the infinite-dimensional parameters was proposed. Our simulation studies showed that the performance of the method was satisfactory even with a very small number of clusters and, also, under a misspecified parametric model for the cause of failure $\pi_{1}(W,\boldsymbol{\gamma}_0)$. In contrast, the previously proposed method by \citet{Bakoyannis20}, which ignores the within-cluster dependence and the informative cluster size, provided biased estimates, underestimated standard errors, and poor coverage probabilities. The analysis of the EA-IeDEA data illustrated that ignoring the within-cluster dependence and the informative cluster size may lead to dubiously significant results in practice. Last but not least, our proposed method can be easily implemented using the \texttt{coxph} function of the R package \texttt{survival}, via a simple data manipulation procedure. This is illustrated in Web Appendix A.

To the best of our knowledge, the issue of clustered competing risks data with missing cause of failure has only been addressed using frailty models by \citet{Lee17}. However, this methodology imposes strong assumptions regarding the within-cluster dependence and the distribution of the random effects, which may be violated in practice. Moreover, this approach does not account for informative cluster size, and does not provide inference about the infinite-dimensional parameters such as the cumulative incidence function. Nevertheless, the covariate-specific cumulative incidence functions are essential for personalized risk prediction in modern medicine. Finally, the method is computationally intensive and provides cluster-specific inference, even though population-averaged inference is more scientifically relevant in many applications including our motivating EA-IeDEA study. The methodology presented in this paper effectively addresses all these limitations and is the first, to the best of our knowledge, rigorous approach for marginal analysis of clustered competing risks data with informative cluster size and missing causes of failure.

Our method adopted a parametric model for the marginal cause of failure probability $\pi_l(\boldsymbol{W}_{i j},\boldsymbol{\gamma}_{0})$, $l=1,\ldots,k$, under a MAR assumption. 
Our simulation studies provided numerical evidence that our regression parameter estimators are robust against some degree of model misspecification. However, the confidence bands had lower coverage rate under a misspecified model for the cause of failure probability. This issue can be alleviated in practice by using flexible parametric models including regression B-splines (with fixed number of internal knots). We also proposed a goodness of fit procedure based on a cumulative residual process to evaluate the model assumption for $\pi_l(\boldsymbol{W}_{i j},\boldsymbol{\gamma}_{0})$, $l=1,\ldots,k$. In our HIV data application, the use of this approach revealed that there was no evidence for a violation of the parametric model assumption.


There are many possible extensions on this work. One may consider nonparametric or semiparametric models for the marginal probability $\pi_l(\boldsymbol{W}_{i j},\boldsymbol{\gamma}_{0})$, or machine learning methods to predict the missing causes of failure. Moreover, extending the methodology for more general clustered and incomplete multi-state data \citep{ Liquet12, Lan17, Bakoyannis19Biometrics} is of interest from both practical and theoretical standpoints.




\section*{Acknowledgements}

Research reported in this publication was supported by the National Institute Of Allergy And Infectious Diseases (NIAID), Eunice Kennedy Shriver National Institute Of Child Health \& Human Development (NICHD), National Institute On Drug Abuse (NIDA), National Cancer Institute (NCI), and the National Institute of Mental Health (NIMH), in accordance with the regulatory requirements of the National Institutes of Health under Award Numbers U01AI069911 and R21AI145662. The content is solely the responsibility of the authors and does not necessarily represent the official views of the National Institutes of Health. This research has also been supported by Lilly Endowment, Inc., through its support for the Indiana University Pervasive Technology Institute, and by the President’s Emergency Plan for AIDS Relief (PEPFAR) through USAID under the terms of Cooperative Agreement No. AID-623-A-12-0001 it is made possible through joint support of the United States Agency for International Development (USAID). The contents of this journal article are the sole responsibility of AMPATH and do not necessarily reflect the views of USAID or the United States Government. \vspace*{-8pt}

\bibliographystyle{chicago}
\bibliography{biblio}

\appendix


\section{R Code}
\label{s:webA}

Our methodology can be easily implemented using standard software that allows for weights. In this Appendix we illustrate the use of the \texttt{coxph} function in the R package \texttt{survival}. Let \texttt{data} be a data set with the variables: cluster id \texttt{clusterid}, observed time \texttt{x}, cause of failure \texttt{c}, missingness indicator \texttt{r}, a covariate of interest \texttt{z}, an auxiliary variable \texttt{a}, and cluster size \texttt{clustersize}.

For illustration purposes, we analyze the cause specific hazard for cause 1 here. The first stage of the analysis is to estimate $\boldsymbol{\gamma}_{0}$ through generalized estimating equations (GEE) for logistic regression using the observations with an observed cause of failure. This can be done using the following code.
\begin{verbatim}
    cause <- 1
    data$include <- 1*(data$r==1 & data$c>0)/data$clustersize
    data$outcome <- 1*(data$c==cause)
    model <- geeglm( outcome ~ x + z + a, family = "binomial", 
        data = data, id = clusterid, weight = include, corstr = "independence")
\end{verbatim}
Then, one needs to calculate the estimated probability of the cause of failure $\pi_1(\boldsymbol{W}_{i j},\hat{\boldsymbol{\gamma}}_{n})$.
\begin{verbatim}
    data$yhat <- predict(model, data, type = "response")
\end{verbatim}
The second stage of the analysis is to maximize the weighted partial pseudolikelihood. This can be implemented with the \texttt{coxph} function using the weight option and some simple data management. The data management steps are used to ``remove" the weights from the risk sets for the observations with a missing cause of failure. 
\begin{verbatim}
    data$d <- data$r*(data$c == cause) + (1-data$r)*(data$yhat > 0)  
    data$weight <- data$r + (1-data$r)*data$yhat
    data$weight <- data$weight + (data$weight == 0)
    dt0 <- data[data$r==0, ]
    dt0$weight <- 1 - dt0$weight
    dt0$d <- 0
    data1 <- rbind(data, dt0)
    data1$weight <- data1$weight/data1$clustersize
\end{verbatim}
Then the point estimates of the regression coefficient can be calculated using the augmented dataset \texttt{data1} as follows. 
\begin{verbatim}
    mod <- coxph(Surv(x, d) ~ z, weight = weight, data = data1)
    beta1 <- coef(mod)
\end{verbatim}
Standard errors of the proposed estimators can be estimated using cluster bootstrap. We plan to develop an R package to implement the proposed close-form variance estimators. The cause-specific baseline cumulative hazard function can be calculated with the \texttt{basehaz} function as follows.
\begin{verbatim}
    H1 <- basehaz(mod, centered = FALSE)
\end{verbatim}

Finally, we can get the estimated baseline cumulative incidence function using equation (1) in the main manuscript, based on the estimated regression coefficients and baseline cumulative hazard functions for all causes of failure. For example, in the case of two causes of failure, the baseline cumulative incidence function can be estimated as follows.
\begin{verbatim}
    Haz1 <- H1$hazard
    Haz2 <- H2$hazard
    S <- exp(- Haz1 - Haz2)
    S.minus <- c(1, S[1: (length(S) - 1)])
    Haz1.minus <- c(0, Haz1[1: (length(Haz1) - 1)])
    CIF1 <- cumsum(S.minus * (Haz1 - Haz1.minus))
\end{verbatim}
Standard errors of the proposed estimators can again be estimated via cluster bootstrap. 

\section{Asymptotic Theory Proofs}
\label{s:webB}

The asymptotic properties of the proposed estimators are justified based on empirical process theory \citep{Kosorok08,vdVaart96}. We use the following standard empirical process notations throughout the Appendix B. For any measurable functions $f:\mathcal{D} \mapsto \mathbb{R}$,
\[
\mathbb{P}_{n} f=\frac{1}{n} \sum_{i=1}^{n} f(\boldsymbol{D}_{i}), \text { and } P f=\int_{\mathcal{D}} f d P=E f.
\]
Also, let $K_0$ denote a generic constant that may change from place to place. In the proofs we will consider an arbitrary cause of failure $l\in\{1,\ldots,k\}$.

\subsection{Proof of Theorem 1}
\label{s:webB1}
To prove the consistency of $\hat{\boldsymbol{\beta}}_{n,l}$ we use the consistency conditions for general Z-estimators \citep{Kosorok08}. In light of condition C5, the expected partial pseudoscore function is
\[
\boldsymbol{G}_{l}(\boldsymbol{\beta})=P\left[\frac{1}{M} \sum_{j=1}^{m_{0}} I(M \geq j)\int_{0}^{\tau}\{\boldsymbol{Z}_{j}-\boldsymbol{E}(t, \boldsymbol{\beta})\} d \tilde{N}_{j l}(t ; \boldsymbol{\gamma}_{0})\right].
\]
Now, by the definition of $\tilde{N}_{ijl}(t;\boldsymbol{\gamma}_0)$ and the assumption of correct specification of the marginal model $\pi_l(\boldsymbol{W}_{ij},\boldsymbol{\gamma}_0)$, it follows that
\[
\boldsymbol{G}_{l}(\boldsymbol{\beta})=P\left[\frac{1}{M} \sum_{j=1}^{m_{0}} I(M \geq j)\int_{0}^{\tau} \{\boldsymbol{Z}_{j}-\boldsymbol{E}(t, \boldsymbol{\beta})\} dN_{jl}(t)\right].
\]
By conditions C5 and C7 we have 
\begin{eqnarray}
\boldsymbol{G}_{l}(\boldsymbol{\beta})&=&
E\left(\frac{1}{M} \sum_{j=1}^{m_{0}} I(M \geq j)E\left[\int_{0}^{\tau} \{\boldsymbol{Z}_{j}-\boldsymbol{E}(t, \boldsymbol{\beta})\} dN_{jl}(t)\Big|M\right]\right) \nonumber \\
&=& E\left(E\left[\int_{0}^{\tau} \{\boldsymbol{Z}_{1}-\boldsymbol{E}(t, \boldsymbol{\beta})\} dN_{1l}(t)\Big|M\right]\frac{1}{M} \sum_{j=1}^{m_{0}} I(M \geq j)\right) \nonumber \\
&=& E\left(E\left[\int_{0}^{\tau} \{\boldsymbol{Z}_{j}-\boldsymbol{E}(t, \boldsymbol{\beta})\} dN_{jl}(t)\Big|M\right]\right) \nonumber \\
&=& E\left[\int_{0}^{\tau} \{\boldsymbol{Z}_{j}-\boldsymbol{E}(t, \boldsymbol{\beta})\} dN_{jl}(t)\right], \label{C7}
\end{eqnarray}
for any $j=1,\ldots,M$. Letting $\Lambda_{0,l}(t)=\int_0^t\lambda_{0,l}(s)ds$, $\boldsymbol{G}_{l}(\boldsymbol{\beta}_{0,l})$ can be expressed as
\begin{eqnarray}
\boldsymbol{G}_{l}(\boldsymbol{\beta}_{0,l})&=& E\left[\int_{0}^{\tau} \{\boldsymbol{Z}_{j}-\boldsymbol{E}(t, \boldsymbol{\beta}_{0,l})\} d\{N_{jl}(t)-Y_{j}(t)\exp(\boldsymbol{\beta}_{0,l}^T\boldsymbol{Z}_{j})d\Lambda_{0,l}(t)\}\right] \nonumber \\
&&+E\left[\int_{0}^{\tau} \{\boldsymbol{Z}_{j}-\boldsymbol{E}(t, \boldsymbol{\beta}_{0,l})\}Y_{j}(t)\exp(\boldsymbol{\beta}_{0,l}^T\boldsymbol{Z}_{j})d\Lambda_{0,l}(t)\right] \label{Gb0}.
\end{eqnarray}
Note that, by the independent censoring assumption conditionally on $\boldsymbol{Z}_{ij}$ and under the assumed marginal propotional cause-specific hazards model, we have
\begin{eqnarray}
E\{N_{jl}(t)|\boldsymbol{Z}_{j}\}&=&P(X_{j}\leq t,C_{j}=l|\boldsymbol{Z}_{j}) \nonumber \\
&=& P(U_{j}\geq T_{j},T_{j}\leq t,C_{j}=l|\boldsymbol{Z}_{j}) \nonumber \\
&=& \int_0^t P(U_{j}\geq s|\boldsymbol{Z}_{j})dF_l(s;\boldsymbol{Z}_{j}) \nonumber \nonumber \\
&=& \int_0^t P(U_{j}\geq s|\boldsymbol{Z}_{j})P(T_{j}>s|\boldsymbol{Z}_{j})\frac{dF_l(s;\boldsymbol{Z}_{j})}{P(T_{j}>s|\boldsymbol{Z}_{j})} \nonumber \\
&=& \int_0^t P(X_{j}\geq s|\boldsymbol{Z}_{j})\exp(\boldsymbol{\beta}_{0,l}^T\boldsymbol{Z}_{j})d\Lambda_{0,l}(s)  \nonumber \\
&=&E\left\{\int_0^t Y_{j}(s)\exp(\boldsymbol{\beta}_{0,l}^T\boldsymbol{Z}_{j})d\Lambda_{0,l}(s)\Big|\boldsymbol{Z}_{j}\right\} \label{ENt}
\end{eqnarray}
for any $j=1,\ldots,M$. Also, by conditions C5 and C7 and similar calculations to those in \eqref{C7}, we have that
\[
\boldsymbol{E}(t, \boldsymbol{\beta}_{0,l})=\frac{E\left\{Y_{j}(t)\exp(\boldsymbol{\beta}^T\boldsymbol{Z}_{j})\boldsymbol{Z}_{j}\right\}}{E\left\{Y_{j}(t)\exp(\boldsymbol{\beta}^T\boldsymbol{Z}_{j})\right\}}.
\]
This fact, equality \eqref{ENt}, and some algebra lead to the conclusion that 
\[
E\left[\int_{0}^{\tau} \{\boldsymbol{Z}_{j}-\boldsymbol{E}(t, \boldsymbol{\beta}_{0,l})\} d\{N_{jl}(t)-Y_{j}(t)\exp(\boldsymbol{\beta}_{0,l}^T\boldsymbol{Z}_{j})d\Lambda_{0,l}(t)\}\right]=\mathbf{0}.
\]
and
\[
E\left[\int_{0}^{\tau} \{\boldsymbol{Z}_{j}-\boldsymbol{E}(t, \boldsymbol{\beta}_{0,l})\}Y_{j}(t)\exp(\boldsymbol{\beta}_{0,l}^T\boldsymbol{Z}_{j})d\Lambda_{0,l}(t)\right]=\mathbf{0}.
\]
Therefore, by \eqref{Gb0}, it follows that $\boldsymbol{G}_{l}(\boldsymbol{\beta}_{0,l})=\mathbf{0}$. Condition C6 implies that $\boldsymbol{\beta}_{0,l}$ is the unique root of $\boldsymbol{G}_{l}(\boldsymbol{\beta})=\mathbf{0}$, $\boldsymbol{\beta}\in\mathcal{B}_l$. To complete the proof of the strong consistency of $\hat{\boldsymbol{\beta}}_{n,l}$, it remains to show that 
\[
\sup_{\boldsymbol{\beta}\in\mathcal{B}_l}\|\boldsymbol{G}_{n, l}(\boldsymbol{\beta}, \hat{\boldsymbol{\gamma}}_{n})-\boldsymbol{G}_{l}(\boldsymbol{\beta})\|\rightarrow_{as*}0.
\]
Using empirical process notation, the empirical version of the pseudoscore function defined in Section 2.2 can be written as
\[
\boldsymbol{G}_{n, l}(\boldsymbol{\beta}, \hat{\boldsymbol{\gamma}}_{n})=\mathbb{P}_{n}\left[\frac{1}{M} \sum_{j=1}^{m_{0}} \int_{0}^{\tau} I(M \geq j)\{\boldsymbol{Z}_{j}-\boldsymbol{E}_{n}(t, \boldsymbol{\beta})\} d \tilde{N}_{j l}(t ; \hat{\boldsymbol{\gamma}}_{n})\right].
\]
The difference between the empirical partial pseudoscore function and expected partial pseudoscore function can be decomposed as
\[
\boldsymbol{G}_{n, l}(\boldsymbol{\beta}, \hat{\boldsymbol{\gamma}}_{n})-\boldsymbol{G}_{l}(\boldsymbol{\beta})=\boldsymbol{A}_{n, l}+\boldsymbol{B}_{n, l}-\boldsymbol{C}_{n, l}(\boldsymbol{\beta})-\boldsymbol{D}_{n, l}(\boldsymbol{\beta})-\boldsymbol{E}_{n, l}(\boldsymbol{\beta}),
\]
where
\[
\boldsymbol{A}_{n, l}=\mathbb{P}_{n}\left[\frac{1}{M} \sum_{j=1}^{m_{0}} \int_{0}^{\tau} I(M \geq j) \boldsymbol{Z}_{j}\{d \tilde{N}_{j l}(t ; \hat{\boldsymbol{\gamma}}_{n})-d \tilde{N}_{j l}(t ; \boldsymbol{\gamma}_{0})\}\right],
\]
\[
\boldsymbol{B}_{n, l}=(\mathbb{P}_{n}-P)\left\{\frac{1}{M} \sum_{j=1}^{m_{0}} \int_{0}^{\tau} I(M \geq j) \boldsymbol{Z}_{j} d \tilde{N}_{j l}(t ; \boldsymbol{\gamma}_{0})\right\},
\]
\[
\boldsymbol{C}_{n, l}(\boldsymbol{\beta})=\mathbb{P}_{n}\left[\frac{1}{M} \sum_{j=1}^{m_{0}} \int_{0}^{\tau} I(M \geq j) \boldsymbol{E}_{n}(t, \boldsymbol{\beta})\{d \tilde{N}_{j l}(t ; \hat{\boldsymbol{\gamma}}_{n})-d \tilde{N}_{j l}(t ; \boldsymbol{\gamma}_{0})\}\right],
\]
\[
\boldsymbol{D}_{n, l}(\boldsymbol{\beta})=\mathbb{P}_{n}\left[\frac{1}{M} \sum_{j=1}^{m_{0}} \int_{0}^{\tau} I(M \geq j)\{\boldsymbol{E}_{n}(t, \boldsymbol{\beta})-\boldsymbol{E}(t, \boldsymbol{\beta})\} d \tilde{N}_{j l}(t ; \boldsymbol{\gamma}_{0})\right],
\]
and
\[
\boldsymbol{E}_{n, l}(\boldsymbol{\beta})=(\mathbb{P}_{n}-P)\left\{\frac{1}{M} \sum_{j=1}^{m_{0}} \int_{0}^{\tau} I(M \geq j) \boldsymbol{E}(t, \boldsymbol{\beta}) d \tilde{N}_{j l}(t ; \boldsymbol{\gamma}_{0})\right\}.
\]
By conditions C3-C5,
\begin{eqnarray*}
\|\boldsymbol{A}_{n, l}\|&=&\left\|\mathbb{P}_{n}\left[\frac{1}{M} \sum_{j=1}^{m_{0}} \int_{0}^{\tau} I(M \geq j) \boldsymbol{Z}_{j}\{d \tilde{N}_{j l}(t ; \hat{\boldsymbol{\gamma}}_{n})-d \tilde{N}_{j l}(t ; \boldsymbol{\gamma}_{0})\}\right]\right\| \\
&=&\left\|\mathbb{P}_{n}\left[\frac{1}{M} \sum_{j=1}^{m_{0}} \int_{0}^{\tau} I(M \geq j) \boldsymbol{Z}_{j}(1-R_{j})\{\pi_{l}(\boldsymbol{W}_{j}, \hat{\boldsymbol{\gamma}}_{n})-\pi_{l}(\boldsymbol{W}_{j}, \boldsymbol{\gamma}_{0})\} d N_{j}(t)\right]\right\| \\
&\leq& K_0\| \hat{\boldsymbol{\gamma}}_{n}-\boldsymbol{\gamma}_{0}\| \left\|\mathbb{P}_{n}\left\{\frac{1}{M} \sum_{j=1}^{m_{0}} \int_{0}^{\tau} I(M \geq j) \boldsymbol{Z}_{j}(1-R_{j}) d N_{j}(t)\right\}\right\| \\
&\rightarrow_{as}& \boldsymbol{0}.
\end{eqnarray*}
By the strong law of large numbers and condition C5,
\begin{eqnarray*}
\boldsymbol{B}_{n, l}&=&(\mathbb{P}_{n}-P)\left\{\frac{1}{M} \sum_{j=1}^{m_{0}} \int_{0}^{\tau} I(M \geq j) \boldsymbol{Z}_{j} d \tilde{N}_{j l}(t ; \boldsymbol{\gamma}_{0})\right\} \rightarrow_{as} \boldsymbol{0}.
\end{eqnarray*}
By conditions C2-C5,
\begin{eqnarray*}
\sup_{\boldsymbol{\beta}\in\mathcal{B}_l}\|\boldsymbol{C}_{n, l}(\boldsymbol{\beta})\|&=&\sup_{\boldsymbol{\beta}\in\mathcal{B}_l}\left\|\mathbb{P}_{n}\left[\frac{1}{M} \sum_{j=1}^{m_{0}} \int_{0}^{\tau} I(M \geq j) \boldsymbol{E}_{n}(t, \boldsymbol{\beta})\{d \tilde{N}_{j l}(t ; \hat{\boldsymbol{\gamma}}_{n})-d \tilde{N}_{j l}(t ; \boldsymbol{\gamma}_{0})\}\right]\right\| \\
&\leq&\sup_{\boldsymbol{\beta}\in\mathcal{B}_l}\left\|\mathbb{P}_{n}\left[\frac{1}{M} \sum_{j=1}^{m_{0}} \int_{0}^{\tau} I(M \geq j) \boldsymbol{E}_{n}(t, \boldsymbol{\beta})\{\pi_{l}(\boldsymbol{W}_{j}, \boldsymbol{\hat{\gamma}}_{n})-\pi_{l}(\boldsymbol{W}_{j}, \boldsymbol{\gamma}_{0})\} d N_{j}(t)\right]\right\| \\
&\leq& K_0\left\|\boldsymbol{\hat{\gamma}}_{n}-\boldsymbol{\gamma}_{0}\right\| \sup_{\boldsymbol{\beta}\in\mathcal{B}_l}\left\|\mathbb{P}_{n}\left\{\frac{1}{M} \sum_{j=1}^{m_{0}} \int_{0}^{\tau} I(M \geq j) \boldsymbol{E}_{n}(t, \boldsymbol{\beta}) d N_{j}(t)\right\}\right\| \\
&\rightarrow_{as^*}& 0.
\end{eqnarray*}
For $\boldsymbol{D}_{n, l}(\boldsymbol{\beta})$, the classes of functions $\{M^{-1} \sum_{j=1}^{m_{0}} I(M \geq j) Y_{j}(t) \exp (\boldsymbol{\beta}^{T} \boldsymbol{Z}_{j}), t \in[0, \tau], \boldsymbol{\beta} \in \mathcal{B}_{l}\}$ and
$\{M^{-1} \sum_{j=1}^{m_{0}} I(M \geq j) Y_{j}(t) \exp (\boldsymbol{\beta}^{T} \boldsymbol{Z}_{j}) \boldsymbol{Z}_{j}, t \in[0, \tau], \boldsymbol{\beta} \in \mathcal{B}_{l}\}$ are Donsker by condition C5 and, thus, also Glivenko-Cantelli. Therefore, using conditions C2 and C5,
\begin{eqnarray*}
\sup_{\boldsymbol{\beta}\in\mathcal{B}_l}\|\boldsymbol{D}_{n, l}(\boldsymbol{\beta})\|&=&\sup_{\boldsymbol{\beta}\in\mathcal{B}_l}\left\|\mathbb{P}_{n}\left[\frac{1}{M} \sum_{j=1}^{m_{0}} \int_{0}^{\tau} I(M \geq j)\{\boldsymbol{E}_{n}(t, \boldsymbol{\beta})-\boldsymbol{E}(t, \boldsymbol{\beta})\} d \tilde{N}_{j l}(t ; \boldsymbol{\gamma}_{0})\right]\right\| \\
&\leq& \sup _{t \in[0, \tau], \boldsymbol{\beta} \in \mathcal{B}_{l}}\|\boldsymbol{E}_{n}(t, \boldsymbol{\beta})-\boldsymbol{E}(t, \boldsymbol{\beta})\| \sup_{\boldsymbol{\beta}\in\mathcal{B}_l}\left\|\mathbb{P}_{n}\left\{\frac{1}{M} \sum_{j=1}^{m_{0}} \int_{0}^{\tau} I(M \geq j) d \tilde{N}_{j i}(t ; \boldsymbol{\gamma}_{0})\right\}\right\| \\
&\leq& K_{0} \sup _{t \in[0, \tau], \boldsymbol{\beta}_{l} \in \mathcal{B}_{l}}\|\boldsymbol{E}_{n}(t, \boldsymbol{\beta})-\boldsymbol{E}(t, \boldsymbol{\beta})\|\\
&\rightarrow_{a s^{*}}& 0.
\end{eqnarray*}
For $\boldsymbol{E}_{n, l}(\boldsymbol{\beta}_{l})$, consider the class of functions
\begin{eqnarray*}
\mathcal{L}_{l}^{(p)}&=&\left\{\frac{1}{M} \sum_{j=1}^{m_{0}} \int_{0}^{\tau} I(M \geq j) \boldsymbol{E}^{(p)}(t, \boldsymbol{\beta}) d \tilde{N}_{j l}(t ; \boldsymbol{\gamma}_{0}), \boldsymbol{\beta} \in \mathcal{B}_{l}\right\}\\
&=&\left\{\frac{1}{M} \sum_{j=1}^{m_{0}}\{R_{j} \Delta_{j t}+(1-R_{j}) \pi_{l}(\boldsymbol{W}_{j}, \boldsymbol{\gamma}_{0})\} \int_{0}^{\tau} I(M \geq j) E^{(p)}(t, \boldsymbol{\beta}) d N_{j}(t), \boldsymbol{\beta} \in \mathcal{B}_{l}\right\},
\end{eqnarray*}
where
\[
E^{(p)}(t, \boldsymbol{\beta})=\frac{E\{\frac{1}{M} \sum_{j=1}^{m_{0}} I(M \geq j) Y_{j}(t) \exp (\boldsymbol{\beta}^{T} \boldsymbol{Z}_{j}) Z_{j}^{(p)}\}}{E\{\frac{1}{M} \sum_{j=1}^{m_{0}} I(M \geq j) Y_{j}(t) \exp (\boldsymbol{\beta}^{T} \boldsymbol{Z}_{j})\}},
\]
with $Z_{j}^{(p)}$ being the $p$th component of $\boldsymbol{Z}_j$. For an arbitrary probability measure $Q$, define the norm $\|f\|_{Q,2}=(\int f^2dQ)^{1/2}$. Now, for any finitely discrete probability measure $Q$ and $\forall \boldsymbol{\beta}_{1}$, $\boldsymbol{\beta}_{2} \in \mathcal{B}_{l}$ and $f_{\boldsymbol{\beta}_{1}}^{(p)}, f_{\boldsymbol{\beta}_{2}}^{(p)} \in \mathcal{L}_{l}^{(p)}$
\begin{eqnarray*}
&&\left\|f_{\boldsymbol{\beta}_{1}}^{(p)}-f_{\boldsymbol{\beta}_{2}}^{(p)}\right\|_{Q, 2}\\
&\leq& \left\| \frac{1}{M} \sum_{j=1}^{m_{0}}\{R_{j} \Delta_{j l}+(1-R_{j}) \pi_{l}(\boldsymbol{W}_{j}, \boldsymbol{\gamma}_{0})\} \int_{0}^{\tau} I(M \geq j)|E^{(p)}(t, \boldsymbol{\beta}_{1})-E^{(p)}(t, \boldsymbol{\beta}_{2})| d N_{j}(t)\right\|_{Q,2}\\
&\leq& \left\| \frac{1}{M} \sum_{j=1}^{m_{0}} \int_{0}^{\tau}|E^{(p)}(t, \boldsymbol{\beta}_{1})-E^{(p)}(t, \boldsymbol{\beta}_{2})| d N_{j}(t)\right\|_{Q, 2}\\
&\leq& K_0\|\boldsymbol{\beta}_{1}-\boldsymbol{\beta}_{2}\|,
\end{eqnarray*}
by the Lipschitz continuity of $E^{(p)}(t, \boldsymbol{\beta})$ and condition C5. Then for $\forall \boldsymbol{\beta} \in \mathcal{B}_{l}$ there exists a $\boldsymbol{\beta}_{i}$, $i=1, \ldots, N(\epsilon, \mathcal{B}_{l},\|\cdot\|)$ such that $\|\boldsymbol{\beta}_{i}-\boldsymbol{\beta} \|<\epsilon$. Therefore, $\forall f_{\boldsymbol{\beta}}^{(p)} \in \mathcal{L}_{l}^{(p)}$ there exists a $f_{\boldsymbol{\beta}_{i}}^{(p)}$ such that
\[
\| f_{\boldsymbol{\beta}_{i}}^{(p)}-f_{\boldsymbol{\beta}}^{(p)} \|_{Q, 2} \leq K_0\| \boldsymbol{\beta}_{i}-\boldsymbol{\beta}\| \equiv \epsilon^{\prime}.
\]
Therefore, $\mathcal{L}_{l}^{(p)}$ can be covered by $N(\epsilon, \mathcal{B}_{l},\|\cdot\|) \ L_{2}(Q) \ \epsilon^{\prime}$-balls centered at $f_{\boldsymbol{\beta}_{i}}^{(p)}$. Thus, the covering number for $\mathcal{L}_{l}^{(p)}$ is $N(\epsilon^{\prime}, \mathcal{L}_{l}^{(p)}, L_{2}(Q)) \leq N(\epsilon, \mathcal{B}_{l},\|\cdot\|)$, where $\mathcal{B}_l$ is a Donsker class as a consequence of condition C2. In addition, using similar arguments to those in page 142 in \citet{Kosorok08}, the class $\mathcal{L}_{l}^{(p)}$ is pointwise measurable. Consequently, $\mathcal{L}_{l}^{(p)}$ is Donsker and, thus, also Glivenko-Cantelli. Therefore, $\sup _{\boldsymbol{\beta} \in \mathcal{B}_{l}}\|\boldsymbol{E}_{n, l}(\boldsymbol{\beta})\| \rightarrow_{a s^{*}} 0$ and, thus, 
\[
\sup _{\boldsymbol{\beta} \in \mathcal{B}_{l}} \| \boldsymbol{G}_{n, l}(\boldsymbol{\beta}, \hat{\boldsymbol{\gamma}}_{n})-\boldsymbol{G}_{l}(\boldsymbol{\beta})|| \rightarrow_{a s^{*}} 0.
\]
This concludes the proof that $\|\hat{\boldsymbol{\beta}}_{n,l}-\boldsymbol{\beta}_{0,l}\| \rightarrow_{a s^{*}} 0$.

Next, we prove the uniform consistency of $\hat{\Lambda}_{1}(t)$. Equality \eqref{ENt}, conditions C5 and C7, and calculations similar to those in \eqref{C7} lead to the conclusion that
\[
\Lambda_{0, l}(t)=\int_{0}^{t} \frac{P\{\frac{1}{M} \sum_{j=1}^{M} d \tilde{N}_{j l}(t ; \boldsymbol{\gamma}_{0})\}}{P\{\frac{1}{M} \sum_{j=1}^{M} Y_{j}(t) \exp (\boldsymbol{\beta}_{0, l}^{T} \boldsymbol{Z}_{j})\}}.
\]
Now, after some algebra, we have that
\begin{eqnarray}
\hat{\Lambda}_{n,l}(t)-\Lambda_{0, l}(t)=A_{n, l}^{*}(t)+B_{n, l}^{*}(t), \label{CH}
\end{eqnarray}
where
\[
A_{n, l}^{*}(t)=\mathbb{P}_{n}\left[\frac{1}{M} \sum_{j=1}^{M}(1-R_{j})\{\pi_{l}(\boldsymbol{W}_{j}, \hat{\boldsymbol{\gamma}}_{n})-\pi_{l}(\boldsymbol{W}_{j}, \boldsymbol{\gamma}_{0})\} \int_{0}^{t} \frac{d N_{j}(t)}{\mathbb{P}_{n}\{\frac{1}{M} \sum_{j=1}^{M} Y_{j}(t) \exp (\hat{\boldsymbol{\beta}}_{n,l}^{T} \boldsymbol{Z}_{j})\}}\right],
\]
\[
B_{n, l}^{*}(t)=\left[\int_{0}^{t} \frac{\mathbb{P}_{n}\{\frac{1}{M} \sum_{j=1}^{M} d \tilde{N}_{j l}(t ; \boldsymbol{\gamma}_{0})\}}{\mathbb{P}_{n}\{\frac{1}{M} \sum_{j=1}^{M} Y_{j}(t) \exp (\hat{\boldsymbol{\beta}}_{n,l}^{T} \boldsymbol{Z}_{j})\}}-\int_{0}^{t} \frac{P\{\frac{1}{M} \sum_{j=1}^{M} d \tilde{N}_{j l}(t ; \boldsymbol{\gamma}_{0})\}}{P\{\frac{1}{M} \sum_{j=1}^{M} Y_{j}(t) \exp (\boldsymbol{\beta}_{0, l}^{T} \boldsymbol{Z}_{j})\}}\right].
\]

By conditions C3, C4 and C5, $\|A_{n, l}^{*}(t)\|_{\infty} \rightarrow_{a s^{*}} 0$. By a similar expansion and arguments in \citet{Kosorok08}, $\|B_{n, l}^{*}(t)\|_{\infty} \rightarrow_{a s^{*}} 0$.
Therefore, $\|\hat{\Lambda}_{n,l}(t)-\Lambda_{0,l}(t)\|_{\infty} \rightarrow_{a s^{*}} 0$.

\subsection{Proof of Theorem 2}
\label{s:webB2}

The estimator $\hat{\boldsymbol{\beta}}_{n, l}$ satisfies 
\begin{eqnarray}
\boldsymbol{0}&=&\sqrt{n} \boldsymbol{G}_{n, l}(\hat{\boldsymbol{\beta}}_{n, l}, \hat{\boldsymbol{\gamma}}_{n}) \nonumber\\
&=&\sqrt{n}\{\boldsymbol{G}_{n, l}(\hat{\boldsymbol{\beta}}_{n, l}, \hat{\boldsymbol{\gamma}}_{n})-\boldsymbol{G}_{n, l}(\hat{\boldsymbol{\beta}}_{n, l}, \boldsymbol{\gamma}_{0})\}+\sqrt{n} \boldsymbol{G}_{n, l}(\hat{\boldsymbol{\beta}}_{n, l}, \boldsymbol{\gamma}_{0}). \label{eqt2}
\end{eqnarray}
The first term in the right side of \eqref{eqt2} can be expressed as
\[
\sqrt{n}\{\boldsymbol{G}_{n, l}(\hat{\boldsymbol{\beta}}_{n, l}, \hat{\boldsymbol{\gamma}}_{n})-\boldsymbol{G}_{n, l}(\hat{\boldsymbol{\beta}}_{n, l}, \boldsymbol{\gamma}_{0})\}=\boldsymbol{A}_{n, l}^{\prime}-\boldsymbol{B}_{n, l}^{\prime}-\boldsymbol{C}_{n, l}^{\prime}+\boldsymbol{D}_{n, l}^{\prime},
\]
where
\[
\boldsymbol{A}_{n, l}^{\prime}=\sqrt{n}(\mathbb{P}_{n}-P)\left[\frac{1}{M} \sum_{j=1}^{m_{0}} I(M \geq j) \boldsymbol{Z}_{j} N_{j}(\tau)(1-R_{j})\{\pi_{l}(\boldsymbol{W}_{j}, \hat{\boldsymbol{\gamma}}_{n})-\pi_{l}(\boldsymbol{W}_{j}, \boldsymbol{\gamma}_{0})\}\right],
\]
\[
\boldsymbol{B}_{n, l}^{\prime}=\sqrt{n}(\mathbb{P}_{n}-P)\left[\frac{1}{M} \sum_{j=1}^{m_{0}} I(M \geq j)(1-R_{j})\{\pi_{l}(\boldsymbol{W}_{j}, \hat{\boldsymbol{\gamma}}_{n})-\pi_{l}(\boldsymbol{W}_{j}, \boldsymbol{\gamma}_{0})\} \int_{0}^{\tau} \boldsymbol{E}(t, \boldsymbol{\beta}_{0, l}) d N_{j}(t)\right],
\]
\[
\boldsymbol{C}_{n, l}^{\prime}=\sqrt{n}(\mathbb{P}_{n}-P)\left[\frac{1}{M} \sum_{j=1}^{m_{0}} \int_{0}^{\tau} I(M \geq j)\{\boldsymbol{E}_{n}(t, \boldsymbol{\beta}_{l})-\boldsymbol{E}(t, \boldsymbol{\beta}_{0, l})\} d\{\tilde{N}_{j l}(t; \hat{\boldsymbol{\gamma}}_{n})-\tilde{N}_{j l}(t ; \boldsymbol{\gamma}_{0})\}\right],
\]
and
\[
\boldsymbol{D}_{n, l}^{\prime}=P\left[\frac{1}{M} \sum_{j=1}^{m_{0}} I(M \geq j)(1-R_{j}) \int_{0}^{\tau}\{\boldsymbol{Z}_{j}-\boldsymbol{E}_{n}(t, \boldsymbol{\beta}_{0, l})\} d N_{j}(t)\dot{\pi}_{l}(\boldsymbol{W}_{j}, \boldsymbol{\gamma}_{0})^{T}\right] \times \sqrt{n}(\hat{\boldsymbol{\gamma}}_{n}-\boldsymbol{\gamma}_{0}).
\]
By conditions C3, C5 and the continuous mapping theorem it follows that $\boldsymbol{A}_{n, l}^{\prime} \rightarrow_{p} \boldsymbol{0}$ and $\boldsymbol{B}_{n, l}^{\prime} \rightarrow_{p} \boldsymbol{0}$.
Using Lemma 4.2 of \citet{Kosorok08}, it follows that $\boldsymbol{C}_{n, l}^{\prime} \rightarrow_{p} \boldsymbol{0}$.
Therefore, the first term in the right side of \eqref{eqt2} is $\sqrt{n}\{\boldsymbol{G}_{n, l}(\hat{\boldsymbol{\beta}}_{n, l}, \hat{\boldsymbol{\gamma}}_{n})-\boldsymbol{G}_{n,l}(\hat{\boldsymbol{\beta}}_{n, l}, \boldsymbol{\gamma}_{0})\}=\boldsymbol{A}_{n,l}^{\prime}-\boldsymbol{B}_{n, l}^{\prime}-\boldsymbol{C}_{n,l}^{\prime}+\boldsymbol{D}_{n, l}^{\prime}=\boldsymbol{D}_{n, l}^{\prime}+o_{p}(1)$. The second term in the right side of \eqref{eqt2} can be expressed as
\[
\sqrt{n} \boldsymbol{G}_{n, l}(\hat{\boldsymbol{\beta}}_{n, l}, \boldsymbol{\gamma}_{0})=\sqrt{n} \boldsymbol{G}_{n, l}(\boldsymbol{\beta}_{0, l}, \boldsymbol{\gamma}_{0})-\boldsymbol{H}_{l}(\boldsymbol{\beta}_{0, l}) \sqrt{n}(\hat{\boldsymbol{\beta}}_{n, l}-\boldsymbol{\beta}_{0, l})+o_{p}(1+\sqrt{n}\|\hat{\boldsymbol{\beta}}_{n, l}-\boldsymbol{\beta}_{0, l}\|).
\]
By condition C6, $\boldsymbol{H}_{l}(\boldsymbol{\beta}_{0, l})$ is invertible and, thus, there exists a constant $K_0>0$ such that for any $\boldsymbol{\beta}_{l} \in \mathcal{B}_{l}$, $\|\boldsymbol{H}_{l}(\boldsymbol{\beta}_{0, l})(\boldsymbol{\beta}_{l}-\boldsymbol{\beta}_{0, l})\| \geq K_0\|\boldsymbol{\beta}_{l}-\boldsymbol{\beta}_{0, l}\|$.
Therefore, by Taylor expansion around $\boldsymbol{\beta}_{0, l}$, $\|\boldsymbol{G}_{n,l}(\boldsymbol{\beta}_{l}, \boldsymbol{\gamma}_{0})-\boldsymbol{G}_{l}(\boldsymbol{\beta}_{0, l}, \boldsymbol{\gamma}_{0})\| \geq K_0\|\boldsymbol{\beta}_{ l}-\boldsymbol{\beta}_{0, l}\|+o_{p}(\|\boldsymbol{\beta}_{l}-\boldsymbol{\beta}_{0, l}\|)$. Now,
\begin{eqnarray*}
&&\sqrt{n}\}\boldsymbol{G}_{n, l}(\hat{\boldsymbol{\beta}}_{n, l}, \boldsymbol{\gamma}_{0})-\boldsymbol{G}_{l}(\boldsymbol{\beta}_{0, l}, \boldsymbol{\gamma}_{0})\}\\
&=&\sqrt{n}(\mathbb{P}_{n}-P)\left[\frac{1}{M} \sum_{j=1}^{m_{0}} \int_{0}^{\tau} I(M \geq j)\{\boldsymbol{Z}_{j}-E(t, \boldsymbol{\beta}_{0, l})\} d \tilde{N}_{j l}(t ; \boldsymbol{\gamma}_{0})\right]\\
&&+o_{p}(1+\sqrt{n}\|\boldsymbol{\beta}_{n, l}-\boldsymbol{\beta}_{0, l}\|)+o_{p}(1)\\
&=&O_{p}(1)+o_{p}(1+\sqrt{n}\|\hat{\boldsymbol{\beta}}_{n, l}-\boldsymbol{\beta}_{0, l}\|)+o_{p}(1).
\end{eqnarray*}
Consequently, $\{K_0+o_{p}(1)\}\sqrt{n}\|\hat{\boldsymbol{\beta}}_{n, l}-\boldsymbol{\beta}_{0, l}\| \leq O_{p}(1)+o_{p}(1+\sqrt{n}\| \hat{\boldsymbol{\beta}}_{n, l}-\boldsymbol{\beta}_{0, l}\|)+o_p(1)$,
and thus $\sqrt{n}\|\hat{\boldsymbol{\beta}}_{n, l}-\boldsymbol{\beta}_{0, l}\|=O_{p}(1)$.
This leads to the conclusion that $\sqrt{n}\boldsymbol{G}_{n,l}(\hat{\boldsymbol{\beta}}_{n, l}, \boldsymbol{\gamma}_{0})=\sqrt{n} \boldsymbol{G}_{n,l}(\boldsymbol{\beta}_{0, l}, \boldsymbol{\gamma}_{0})-\boldsymbol{H}_{l}(\boldsymbol{\beta}_{0, l}) \sqrt{n}(\hat{\boldsymbol{\beta}}_{n, l}-\boldsymbol{\beta}_{0, l})+o_{p}(1)$. Recalling that $\tilde{M}_{j l}(t; \boldsymbol{\beta}_{0, l}, \boldsymbol{\gamma}_{0})=\tilde{N}_{j l}(t; \boldsymbol{\gamma}_{0})-\int_{0}^{t}I(X_{j} \geq u) \exp (\boldsymbol{\beta}_{0, l}^{T} \boldsymbol{Z}_{j}) d \Lambda_{0, l}(u)$ we have
\[
\sqrt{n} \boldsymbol{G}_{n, l}(\boldsymbol{\beta}_{0, l}, \boldsymbol{\gamma}_{0})=\sqrt{n} \mathbb{P}_{n}\left[\frac{1}{M} \sum_{j=1}^{m_{0}} \int_{0}^{\tau} I(M \geq j)\left\{\boldsymbol{Z}_{j}-\boldsymbol{E}(t, \boldsymbol{\beta}_{0, l})\right\} d \tilde{M}_{j l}(t ; \boldsymbol{\beta}_{0 t}, \boldsymbol{\gamma}_{0})\right]+o_{p}(1).
\]
Taking all the pieces together we have that
\begin{eqnarray*}
\boldsymbol{0}&=&\sqrt{n}\boldsymbol{G}_{n,l}(\hat{\boldsymbol{\beta}}_{n, l}, \hat{\boldsymbol{\gamma}}_{n})\\
&=&\sqrt{n}\{\boldsymbol{G}_{n, l}(\hat{\boldsymbol{\beta}}_{n, l}, \hat{\boldsymbol{\gamma}}_{n})-\boldsymbol{G}_{n, l}(\hat{\boldsymbol{\beta}}_{n, l}, \boldsymbol{\gamma}_{0})\}+\sqrt{n} \boldsymbol{G}_{n, l}(\hat{\boldsymbol{\beta}}_{n, l}, \boldsymbol{\gamma}_{0})\\
&=&P\left\{\frac{1}{M} \sum_{j=1}^{m_{0}} I(M \geq j)(1-R_{j}) \int_{0}^{\tau}[\boldsymbol{Z}_{j}-\boldsymbol{E}_{n}(t, \boldsymbol{\beta}_{0, l})] d N_{j}(t)\dot{\pi}_{l}(\boldsymbol{W}_{j}, \boldsymbol{\gamma}_{0})^{T}\right\} \times \sqrt{n}(\hat{\boldsymbol{\gamma}}_{n}-\boldsymbol{\gamma}_{0})\\
&&+\sqrt{n} \mathbb{P}_{n}\left\{\frac{1}{M} \sum_{j=1}^{m_{0}} \int_{0}^{\tau} I(M \geq j)\left\{\boldsymbol{Z}_{j}-\boldsymbol{E}(t, \boldsymbol{\beta}_{0, l})\right\} d \tilde{M}_{j l}(t ; \boldsymbol{\beta}_{0 t}, \boldsymbol{\gamma}_{0})\right\}\\
&&-\boldsymbol{H}_{l}(\boldsymbol{\beta}_{0, l}) \sqrt{n}(\hat{\boldsymbol{\beta}}_{n, l}-\boldsymbol{\beta}_{0, l})+o_{p}(1).
\end{eqnarray*}
Rearranging the terms and according to conditions C4 and C6 leads to
\begin{eqnarray*}
\sqrt{n}(\hat{\boldsymbol{\beta}}_{n, l}-\boldsymbol{\beta}_{0, l})=\frac{1}{\sqrt{n}} \sum_{i=1}^{n}\left\{\frac{1}{M_{i}} \sum_{j=1}^{M_{i}}(\boldsymbol{\psi}_{i j l}+\boldsymbol{R}_{l}\boldsymbol{\omega}_{i j})\right\}+o_{p}(1),
\end{eqnarray*}
where
\[
\boldsymbol{\psi}_{i j l}={\boldsymbol{H}}_{l}^{-1}(\boldsymbol{\beta}_{0,l})\int_0^{\tau}\{{\boldsymbol{Z}}_{i j}-\boldsymbol{E}(t,\boldsymbol{\beta}_{0,l})\}d\tilde{M}_{i j l}(t;\boldsymbol{\beta}_{0,l},\boldsymbol{\gamma}_0), 
\]
and
\[
R_{l}=\boldsymbol{H}_{l}^{-1}(\boldsymbol{\beta}_{0, l}) E\left[\frac{1}{M} \sum_{j=1}^{M} (1-R_{j})\int_{0}^{\tau}\{\boldsymbol{Z}_{j}-\boldsymbol{E}(t, \boldsymbol{\beta}_{0, l})\} d N_{j}(t)\dot{\pi}_{l}(\boldsymbol{W}_{j}, \boldsymbol{\gamma}_{0})^{T}\right].
\]

\subsection{Proof of Theorem 3}
\label{s:webB3}

By Taylor expansion and the consistency of $\hat{\boldsymbol{\beta}}_{n,l}$ and $\hat{\boldsymbol{\gamma}}_{n}$, the first term in the right side of expansion \eqref{CH} can be written as 
\begin{eqnarray*}
A_{n, l}^{*}(t)&=&\mathbb{P}_{n}\left[\frac{1}{M} \sum_{j=1}^{M}(1-R_{j})\{\pi_{l}(\boldsymbol{W}_{j}, \hat{\boldsymbol{\gamma}}_{n})-\pi_{l}(\boldsymbol{W}_{j}, \boldsymbol{\gamma}_{0})\} \int_{0}^{t} \frac{d N_{j}(t)}{\mathbb{P}_{n}\{\frac{1}{M} \sum_{j=1}^{M} Y_{j}(t) \exp (\hat{\boldsymbol{\beta}}_{n,l}^{T} \boldsymbol{Z}_{j})\}}\right]\\
&=&\frac{1}{\sqrt{n}} \sum_{i=1}^{n}\left\{ \frac{1}{M_{i}} \sum_{j=1}^{M_{i}} \boldsymbol{R}_{l}^{*}(t) \boldsymbol{\omega}_{i j}\right\}+o_{p}(n^{-1/2}),
\end{eqnarray*}
where
\[
{\boldsymbol{R}}_j^{\star}(t)=E\left[\frac{1}{M} \sum_{j=1}^{M} (1-R_{j})\dot{\pi}_j({\boldsymbol{W}_{j}},\boldsymbol{\gamma}_0)\int_0^t\frac{dN_{j}(s)}{E\{\frac{1}{M} \sum_{j=1}^{M} Y_{j}(s) \exp (\boldsymbol{\beta}_{0,l}^{T} \boldsymbol{Z}_{j})\}}\right]^T.
\]
By similar analysis to that provided in Page 57 of \citet{Kosorok08}, the second term in \eqref{CH} can be written as
\begin{eqnarray*}
B_{n, l}^{*}(t)&=&\left[\int_{0}^{t} \frac{\mathbb{P}_{n}\{\frac{1}{M} \sum_{j=1}^{M} d \tilde{N}_{j l}(t ; \boldsymbol{\gamma}_{0})\}}{\mathbb{P}_{n}\{\frac{1}{M} \sum_{j=1}^{M} Y_{j}(t) \exp (\hat{\boldsymbol{\beta}}_{n,l}^{T} \boldsymbol{Z}_{j})\}}-\int_{0}^{t} \frac{P\{\frac{1}{M} \sum_{j=1}^{M} d \tilde{N}_{j l}(t ; \boldsymbol{\gamma}_{0})\}}{P\{\frac{1}{M} \sum_{j=1}^{M} Y_{j}(t) \exp (\boldsymbol{\beta}_{0, l}^{T} \boldsymbol{Z}_{j})\}}\right]\\
&=&\frac{1}{\sqrt{n}} \sum_{i=1}^{n}\left\{ \frac{1}{M_{i}} \sum_{j=1}^{M_{i}} \phi_{i j l}(t)\right\}+o_{p}(n^{-1/2}),
\end{eqnarray*}
where
\[
\phi_{ijl}(t)=\int_0^t\frac{d\tilde{M}_{ijl}(s;\boldsymbol{\beta}_{0,j},\boldsymbol{\gamma}_0)}{E\{\frac{1}{M} \sum_{j=1}^{M} Y_{j}(s) \exp (\boldsymbol{\beta}_{0,l}^{T} \boldsymbol{Z}_{j})\}}-(\boldsymbol{\psi}_{ijl}+{\boldsymbol{R}}_l\boldsymbol{\omega}_{ij})^T\int_0^tE(s,\boldsymbol{\beta}_{0,l})d\Lambda_{0,l}(s).
\]
Therefore,
\[
\sqrt{n}\left\{\hat{\Lambda}_{n, l}(t)-\Lambda_{0, l}(t)\right\}=\frac{1}{\sqrt{n}} \sum_{i=1}^{n}\left[\frac{1}{M_{i}} \sum_{j=1}^{M_{i}} \{\phi_{i j l}(t)+\boldsymbol{R}_{l}^{*}(t) \boldsymbol{\omega}_{i j}\}\right]+o_{p}(1).
\]
By conditions C1, C4, and C5, and lemmas 1 and 2 in the supporting information of \citet{Bakoyannis19Biometrics}, the class of functions
\[
\left[\frac{1}{M} \sum_{j=1}^{M} \{\phi_{j l}(t)+\boldsymbol{R}_{l}^{*}(t) \boldsymbol{\omega}_{j}\}:t\in[0,\tau]\right]
\]
is Donsker. We now show that conditional on the data, the estimated multiplier process
\[
\hat{W}_{n, l}(\cdot)=\frac{1}{\sqrt{n}} \sum_{i=1}^{n}\left[ \frac{1}{M_{i}} \sum_{j=1}^{M_{i}} \{\hat{\phi}_{i j l}(\cdot)+\hat{\boldsymbol{R}}_{l}^{*}(\cdot) \hat{\boldsymbol{\omega}}_{i j}\}\right]\xi_{i}
\]
converges weakly to the same limiting process as $W_{n, l}(\cdot)=\sqrt{n}\{\hat{\Lambda}_{n, l}(\cdot)-\Lambda_{0, l}(\cdot)\}$. Define
\[
\tilde{W}_{n, l}(\cdot)=\frac{1}{\sqrt{n}} \sum_{i=1}^{n}\left[ \frac{1}{M_{i}} \sum_{j=1}^{M_{i}} \{\phi_{i j l}(\cdot)+\boldsymbol{R}_{l}^{*}(\cdot) \boldsymbol{\omega}_{i j}\}\right]\xi_{i}.
\]
By the Donsker property of the class of influence functions and the conditional multiplier central limit theorem \citep{vdVaart96}, $\tilde{W}_{n, l}(\cdot)$ converges weakly, conditionally on the data, to the same limiting process as $W_{n, l}(\cdot)$. To complete the proof, we need to show
\[
\|\hat{W}_{n, l}(t)-\tilde{W}_{n, l}(t)\|_{\infty}=o_{p}(1),
\]
unconditionally. After some algebra
\[
\|\hat{W}_{n, l}(t)-\tilde{W}_{n, l}(t)\|_{\infty} \leq A_{n,l}^{\prime\prime}+B_{n,l}^{\prime\prime}+C_{n,l}^{\prime\prime},
\]
where
\[
A_{n,l}^{\prime\prime}=\left\|\frac{1}{\sqrt{n}} \sum_{i=1}^{n}\left[\frac{1}{M_{i}} \sum_{j=1}^{M_{i}}\left\{\hat{\phi}_{i j l}(t)-\phi_{i j l}(t)\right\}\right] \xi_{i}\right\|_{\infty},
\]
\[
B_{n,l}^{\prime\prime}=\sup _{t \in[0, \tau]}\left\|\hat{\boldsymbol{R}}_{l}^{*}(t)-\boldsymbol{R}_{l}^{*}(t)\right\| \times\left(\left\|\frac{1}{\sqrt{n}} \sum_{i=1}^{n}\left\{\frac{1}{M_{i}} \sum_{j=1}^{M_{i}}(\hat{\boldsymbol{\omega}}_{i j}-\boldsymbol{\omega}_{i j})\right\} \xi_{i}\right\|+\left\|\frac{1}{\sqrt{n}} \sum_{i=1}^{n}\left(\frac{1}{M_{i}} \sum_{j=1}^{M_{i}} \boldsymbol{\omega}_{i j}\right) \xi_{i}\right\|\right),
\]
and
\[
C_{n,l}^{\prime\prime}=\sup _{t \in[0, \tau]}\left\|\boldsymbol{R}_{l}^{*}(t)\right\| \times\left\|\frac{1}{\sqrt{n}} \sum_{i=1}^{n}\left\{\frac{1}{M_{i}} \sum_{j=1}^{M_{i}}(\hat{\boldsymbol{\omega}}_{i j}-\boldsymbol{\omega}_{i j})\right\} \xi_{i}\right\|.
\]
Using the same arguments to those used in the proof of Theorem 4 in \citet{Spiekerman98} and regularity conditions C3 and C4, $A_{n,l}^{\prime\prime}=o_p(1)$. 

For $B_{n,l}^{\prime\prime}$, using the same arguments to those used in the proof of Lemma A.3 in \citet{Spiekerman98} leads to the conclusion that
\[
\left\|\frac{1}{\sqrt{n}} \sum_{i=1}^{n}\left\{\frac{1}{M_{i}} \sum_{j=1}^{M_{i}}(\hat{\boldsymbol{\omega}}_{i j}-\boldsymbol{\omega}_{i j})\right\} \xi_{i}\right\|=o_p(1).
\]
Next, by condition C4 and the central limit theorem
\[
\left\|\frac{1}{\sqrt{n}} \sum_{i=1}^{n}\left(\frac{1}{M_{i}} \sum_{j=1}^{M_{i}} \boldsymbol{\omega}_{i j}\right) \xi_{i}\right\|=O_p(1).
\]
Also, after some algebra, we have that
\[
\sup _{t \in[0, \tau]}\left\|\hat{\boldsymbol{R}}_{l}^{*}(t)-\boldsymbol{R}_{l}^{*}(t)\right\|\leq A_{n,l}^{\prime\prime\prime}+B_{n,l}^{\prime\prime\prime}+C_{n,l}^{\prime\prime\prime},
\]
where
\[
A_{n,l}^{\prime\prime\prime}=\sup _{t \in[0, \tau]}\left\|\mathbb{P}_{n}\left[\frac{1}{M} \sum_{j=1}^{M}\left\{\dot{\pi}_{l}\left(\boldsymbol{W}_{j}, \hat{\boldsymbol{\gamma}}_{n}\right)-\dot{\pi}_{l}\left(\boldsymbol{W}_{j}, \boldsymbol{\gamma}_{0}\right)\right\} \int_{0}^{t} \frac{d N_{j}(s)}{\mathbb{P}_{n}\left\{\frac{1}{M} \sum_{j=1}^{M} Y_{j}(s) \exp \left(\hat{\boldsymbol{\beta}}_{n, l}^{T} \boldsymbol{Z}_{j}\right)\right\}}\right]\right\|,
\]
\[
B_{n,l}^{\prime\prime\prime}=\sup _{t \in[0, \tau]}\left\|\mathbb{P}_{n} \frac{1}{M} \sum_{j=1}^{M} \dot{\pi}_{l}\left(\boldsymbol{W}_{j}, \boldsymbol{\gamma}_{0}\right) \int_{0}^{t}[\frac{1}{\mathbb{P}_{n}\{\frac{1}{M} \sum_{j=1}^{M} Y_{j}(s) e^{\hat{\boldsymbol{\beta}}_{n, l}^{T} \boldsymbol{Z}_{j}}\}}-\frac{1}{P\{\frac{1}{M} \sum_{j=1}^{M} Y_{j}(s) e^{\boldsymbol{\beta}_{0,l}^{T} \boldsymbol{Z}_{j}}\}}] d N_{j}(s)\right\|,
\]
and
\[
C_{n,l}^{\prime\prime\prime}=\sup _{t \in[0, \tau]}\left\|\left(\mathbb{P}_{n}-P\right)\left[\frac{1}{M} \sum_{j=1}^{M} \dot{\pi}_{l}\left(\boldsymbol{W}_{j}, \boldsymbol{\gamma}_{0}\right) \int_{0}^{t} \frac{1}{P\left\{\frac{1}{M} \sum_{j=1}^{M} Y_{j}(s) \exp \left(\boldsymbol{\beta}_{0,l}^{T} \boldsymbol{Z}_{j}\right)\right\}} d N_{j}(s)\right]\right\|.
\]
By conditions C3, C4, C5, and the continuous mapping theorem,
\[
\max _{i j}\left\|\dot{\pi}_{l}\left(\boldsymbol{W}_{i j}, \hat{\gamma}_{n}\right)-\dot{\pi}_{l}\left(\boldsymbol{W}_{i j}, \boldsymbol{\gamma}_{0}\right)\right\|=o_{a s}(1).
\]
By Theorem 2 and conditions C1, C2, and C5,
\[
\left\|\frac{1}{\mathbb{P}_{n}\left\{\frac{1}{M} \sum_{j=1}^{M} Y_{j}(s) \exp (\hat{\boldsymbol{\beta}}_{n, l}^{T} \boldsymbol{Z}_{j})\right\}}\right\|_{\infty}=\left\|\frac{1}{P\left\{\frac{1}{M} \sum_{j=1}^{M} Y_{j}(s) \exp (\boldsymbol{\beta}_{0,l}^{T} \boldsymbol{Z}_{j})\right\}+o_{a s^{*}}(1)}\right\|=O_{a s^{*}}(1).
\]
Therefore, $A_{n,l}^{\prime\prime\prime}=o_{a s^{*}}(1)$. Next, by conditions C3 and C5, $\max _{i j}\left\|\dot{\pi}_{l}\left(\boldsymbol{W}_{i j}, \boldsymbol{\gamma}_{0}\right)\right\|=O_{a s}(1)$. Also, by conditions C2, C5, and the Donsker property of $\left\{Y_{j}(t): t \in[0, \tau]\right\}$,
\[
\left\|\frac{1}{\mathbb{P}_{n}\left\{\frac{1}{M} \sum_{j=1}^{M} Y_{j}(s) \exp \left(\hat{\boldsymbol{\beta}}_{n, l}^{T} \boldsymbol{Z}_{j}\right)\right\}}-\frac{1}{P\left\{\frac{1}{M} \sum_{j=1}^{M} Y_{j}(s) \exp \left(\boldsymbol{\beta}_{0,l}^{T} \boldsymbol{Z}_{j}\right)\right\}}\right\|_{\infty}=o_{a s ^{*}}(1),
\]
and, thus, $B_{n,l}^{\prime\prime\prime}=o_{a s^{*}}(1)$. For $C_{n,l}^{\prime\prime\prime}$, consider the classes of functions $\mathcal{F}=\{N_{j}(s): t \in[0, \tau]\}$ and
\[
\mathcal{L}_{l, 1}=\left\{f_{t, l}=\frac{1}{M} \sum_{j=1}^{m_{0}} I(M \geq j) \dot{\pi}_{l}(\boldsymbol{W}_{j}, \boldsymbol{\gamma}_{0}) \int_{0}^{t} \frac{1}{P\{\frac{1}{M} \sum_{j=1}^{M} Y_{j}(s) e^ {\boldsymbol{\beta}_{0,l}^{T} \boldsymbol{Z}_{j}}\}} d N_{j}(s), t \in[0, \tau]\right\}.
\]
For any finitely discrete probability measure $Q$ and any $t_1$, $t_2 \in [0,\tau]$ we have that
\begin{eqnarray*}
\| f_{t_{1}, l}-f_{t_{2}, l}\|_{Q, 2} &\leq& \left\| \frac{1}{M} \sum_{j=1}^{m_{0}} I(M \geq j) \dot{\pi}_{l}(\boldsymbol{W}_{j}, \boldsymbol{\gamma}_{0}) \int_{t_{1}}^{t_{2}} \frac{d N_{j}(s)}{P\{\frac{1}{M} \sum_{j=1}^{M} Y_{j}(s) \exp (\boldsymbol{\beta}_{0,l}^{T} \boldsymbol{Z}_{j})\}}\right\|_{Q, 2}\\
&\leq& K_0\| N_{j}(t_{2})-N_{j}(t_{1})\|_{Q, 2}.
\end{eqnarray*}
Therefore, $\mathcal{L}_{l, 1}$ can be covered by $N\left(\epsilon, \mathcal{F}, L_{2}(Q)\right)\ L_{2}(Q)\ \epsilon^{\prime}$ -balls centered at $f_{t_{i}, l}$, where $\mathcal{F}$ is a Donsker class by lemma 4.1 in \citet{Kosorok08}. In addition, using similar arguments to those in page 142 in \citet{Kosorok08}, the class $\mathcal{L}_{l, 1}$ is pointwise measurable. Consequently, the class $\mathcal{L}_{l, 1}$ is Donsker and, thus, also Glivenko-Cantelli, which leads to the conclusion that $C_{n,l}^{\prime\prime\prime}=o_{a s^{*}}(1)$. Therefore, $\sup _{t \in[0, \tau]}\|\hat{\boldsymbol{R}}_{l}^{*}(t)-\boldsymbol{R}_{l}^{*}(t)\|=o_{p}(1)$ and, thus. $B_{n, l}^{\prime \prime}=o_{p}(1)$. Similar arguments lead to the conclusion that $C_{n, l}^{\prime \prime}=o_{p}(1)$. Thus, 
\[
\left\|\hat{W}_{n, l}(t)-\tilde{W}_{n, l}(t)\right\|_{\infty}=o_{p}(1),
\]
which completes the proof of the last statement in Theorem 3.

\subsection{Proof of Theorem 4}
\label{s:webB4}

It is easy to show that
\begin{eqnarray*}
\tilde{W}_{n,l}(t ; \boldsymbol{z}_{0})&=&\sqrt{n}\{\hat{\Lambda}_{n,l}(t; \boldsymbol{z}_{0})-\Lambda_{0, l}(t; \boldsymbol{z}_{0})\}\\
&=&\sqrt{n}\{\hat{\Lambda}_{n,l}(t) \exp (\hat{\boldsymbol{\beta}}_{n,l}^{T} \boldsymbol{z}_{0})-\Lambda_{0, l}(t) \exp (\boldsymbol{\beta}_{0,l}^{T} \boldsymbol{z}_{0})\}\\
&=&\sqrt{n}\{\hat{\Lambda}_{n, l}(t)-\Lambda_{0, l}(t)+\boldsymbol{z}_{0}^{T}(\hat{\boldsymbol{\beta}}_{n,l}-\boldsymbol{\beta}_{0, l}) \Lambda_{0, l}(t)\} \exp (\boldsymbol{\beta}_{0, l}^{T} \boldsymbol{z}_{0})+o_{p}(1)\\
&=&\frac{1}{\sqrt{n}} \sum_{i=1}^{n}\frac{1}{M_{i}} \sum_{j=1}^{M_{i}}\{\boldsymbol{z}_{0}^{T}(\boldsymbol{\psi}_{i j l}+\boldsymbol{R}_{l}\boldsymbol{\omega}_{i j}) \Lambda_{0, l}(t)+\phi_{i j l}(t)+\boldsymbol{R}_{l}^{*}(t) \boldsymbol{\omega}_{i j}\} \exp (\boldsymbol{\beta}_{0, l}^{T} \boldsymbol{z}_{0})+o_{p}(1)\\
&=&\frac{1}{\sqrt{n}} \sum_{i=1}^{n} \frac{1}{M_{i}} \sum_{j=1}^{M_{i}} \phi_{i j l}^{\Lambda}(t ; \boldsymbol{z}_{0})+o_{p}(1).
\end{eqnarray*}
Similarly to the decomposition in \citet{Cheng98},
\begin{eqnarray*}
\sqrt{n}\left\{\hat{F}_{n,l}(t ; \boldsymbol{z}_{0})-F_{0,l}(t ; \boldsymbol{z}_{0})\right\}=\frac{1}{\sqrt{n}} \sum_{i=1}^{n} \left\{ \frac{1}{M_{i}} \sum_{j=1}^{M_{i}}\phi_{i j l}^{F}(t ; \boldsymbol{z}_{0})\right\}+o_{p}(1),
\end{eqnarray*}
where
\begin{eqnarray*}
& &\phi_{ijl}^F(t;{\boldsymbol{z}}_0)=\int_0^t\exp\left\{-\sum_{l=1}^k\Lambda_{0,l}(s-;{\boldsymbol{z}}_0)\right\}d\phi^{\Lambda}_{ijl}(s;{\boldsymbol{z}}_0)\\
& &\hspace{25mm}-\int_0^t\left\{\sum_{l=1}^k\phi^{\Lambda}_{ijl}(s-;{\boldsymbol{z}}_0)\right\}\exp\left\{-\sum_{l=1}^k\Lambda_{0,l}(s-;{\boldsymbol{z}}_0)\right\}d\Lambda_{0,l}(s;{\boldsymbol{z}}_0),
\end{eqnarray*}
and
$\phi_{ijl}^{\Lambda}(t;{\boldsymbol{z}}_0)=\{{\boldsymbol{z}}_0^T(\boldsymbol{\psi}_{ijl}+{\boldsymbol{R}}_l\boldsymbol{\omega}_{ij})\Lambda_{0,l}(t)+\phi_{ijl}(t)+{\boldsymbol{R}}_l^{\star}(t)\boldsymbol{\omega}_{ij}\}\exp(\boldsymbol{\beta}_{0,l}^T{\boldsymbol{z}}_0)$. The class of functions $\{\phi_{jl}^F(t;\boldsymbol{z}_0:t\in[0,\tau]\}$ is Donsker by conditions C1, C4, and C5, lemmas 1 and 2 in the supporting information of \citet{Bakoyannis19Biometrics}, and corollary 9.32 in \citet{Kosorok08}.

To conclude the proof of Theorem 4, we show that, conditionally on the data, 
\[
\hat{W}_{n, l}^{F}(\cdot ; \boldsymbol{z}_{0})=\frac{1}{\sqrt{n}} \sum_{i=1}^{n} \frac{1}{M_{i}} \sum_{j=1}^{M_{i}} \hat{\phi}_{i j l}^{F}(\cdot ; \boldsymbol{z}_{0}) \xi_{i}
\]
converges weakly to the same limiting process as $W_{n, l}^{F}(\cdot ; \boldsymbol{z}_{0})=\sqrt{n}\{\hat{F}_{n,l}(\cdot ; \boldsymbol{z}_{0})-F_{0,l}(\cdot ; \boldsymbol{z}_{0})\}$. Now, define 
\[
\tilde{W}_{n, l}^{F}(\cdot ; \boldsymbol{z}_{0})=\frac{1}{\sqrt{n}} \sum_{i=1}^{n} \frac{1}{M_{i}} \sum_{j=1}^{M_{i}} \phi_{i j l}^{F}(\cdot ; \boldsymbol{z}_{0}) \xi_{i}. 
\]
By the Donsker property of the class of influence functions and the conditional multiplier central limit theorem \citep{vdVaart96}, $\tilde{W}_{n, l}^{F}(\cdot ; \boldsymbol{z}_{0})$ converges weakly, conditionally on the data, to the same limiting process as $W_{n, l}^{F}(\cdot ; \boldsymbol{z}_{0})$. To complete the proof, we need to show that
\[
\|\hat{W}_{n, l}^{F}(t ; \boldsymbol{z}_{0})-\tilde{W}_{n, l}^{F}(t ; \boldsymbol{z}_{0})\|_{\infty}=o_{p}(1),
\]
unconditionally. Some algebra leads to the following bound 
\[
\|\hat{W}_{n, l}^{F}(t ; \boldsymbol{z}_{0})-\tilde{W}_{n, l}^{F}(t ; \boldsymbol{z}_{0})\|_{\infty} \leq A_{n,l}^{\prime\prime\prime\prime}+B_{n,l}^{\prime\prime\prime\prime}+C_{n,l}^{\prime\prime\prime\prime}+D_{n,l}^{\prime\prime\prime\prime}+E_{n,l}^{\prime\prime\prime\prime}+F_{n,l}^{\prime\prime\prime\prime},
\]
where
\[
A_{n,l}^{\prime\prime\prime\prime}=\left\|\int_{0}^{t} \exp \left\{-\sum_{l=1}^{k} \hat{\Lambda}_{n, l}\left(s-; \boldsymbol{z}_{0}\right)\right\} d\left[\sqrt{n} \mathbb{P}_{n} \frac{1}{M} \sum_{j=1}^{M}\left\{\hat{\phi}_{j l}^{A}\left(s ; \boldsymbol{z}_{0}\right)-\phi_{j l}^{A}\left(s ; \boldsymbol{z}_{0}\right)\right\} \xi\right]\right\|_{\infty},
\]
\[
B_{n,l}^{\prime\prime\prime\prime}=\left\|\int_{0}^{t} \exp \left\{-\sum_{l=1}^{k} \hat{\Lambda}_{n, l}\left(s-; \boldsymbol{z}_{0}\right)\right\}-\exp \left\{-\sum_{l=1}^{k} \Lambda_{0, l}\left(s-; \boldsymbol{z}_{0}\right)\right\} d\left\{\sqrt{n} \mathbb{P}_{n} \frac{1}{M} \sum_{j=1}^{M} \phi_{j l}^{\Lambda}\left(s ; \boldsymbol{z}_{0}\right) \xi\right\}\right\|_{\infty},
\]
\begin{eqnarray*}
C_{n, j}^{\prime \prime \prime \prime}=& \Big\| \sum_{l=1}^{k} \int_{0}^{t}\left[\mathbb{P}_{n}\frac{1}{M} \sum_{j=1}^{M}\left\{\hat{\phi}_{j l}^{A}\left(s- ; \boldsymbol{z}_{0}\right)-\phi_{j l}^{A}\left(s- ; \boldsymbol{z}_{0}\right)\right\} \xi \right] \exp \left\{-\sum_{l=1}^{k} \hat{\Lambda}_{n, l}\left(s-; \boldsymbol{z}_{0}\right)\right\} \nonumber\\
& \times d\left[\sqrt{n}\left\{\hat{\Lambda}_{n, l}\left(s ; \boldsymbol{z}_{0}\right)-A_{0, l}\left(s ; \boldsymbol{z}_{0}\right)\right\}\right] \Big\|_{\infty},
\end{eqnarray*}
\begin{eqnarray*}
D_{n, j}^{\prime \prime \prime \prime}=& \Big\| \sum_{l=1}^{k} \int_{0}^{t}\left\{\mathbb{P}_{n}\frac{1}{M} \sum_{j=1}^{M}\phi_{j l}^{A}\left(s- ; \boldsymbol{z}_{0}\right) \xi\right\}\left[\exp \{-\sum_{l=1}^{k} \hat{\Lambda}_{n, l}(s-; \boldsymbol{z}_{0})\}-\exp \{-\sum_{l=1}^{k} \Lambda_{n, l}(s-; \boldsymbol{z}_{0})\}\right] \\
& \times d\left[\sqrt{n}\{\hat{\Lambda}_{n, l}(s ; \boldsymbol{z}_{0})-\Lambda_{0, l}(s ; \boldsymbol{z}_{0})\}\right] \Big\|_{\infty},
\end{eqnarray*}
\[
E_{n,l}^{\prime\prime\prime\prime}=\left\|\sum_{l=1}^{k} \int_{0}^{t}\left[\sqrt{n} \mathbb{P}_{n}\frac{1}{M} \sum_{j=1}^{M}\left\{\hat{\phi}_{j l}^{A}\left(s- ; \boldsymbol{z}_{0}\right)-\phi_{j l}^{A}\left(s- ; \boldsymbol{z}_{0}\right)\right\} \xi\right] d \Lambda_{0, l}\left(s ; \boldsymbol{z}_{0}\right)\right\|_{\infty},
\]
\[
F_{n,l}^{\prime\prime\prime\prime}=O_{p}(1)\left|\sqrt{n} \mathbb{P}_{n} \xi\right|\left\|\int_{0}^{t}[\exp \{-\sum_{l=1}^{k} \hat{\Lambda}_{n, l}\left(s-; \boldsymbol{z}_{0}\right)\}-\exp \{-\sum_{l=1}^{k} \Lambda_{0, l}\left(s-; \boldsymbol{z}_{0}\right)\}] d \Lambda_{0, l}\left(s ; \boldsymbol{z}_{0}\right)\right\|_{\infty}.
\]

By integration by parts, Theorem 3, Lemma A.3 in \citet{Spiekerman98} and the boundedness conditions, $A_{n,l}^{\prime\prime\prime\prime}=o_p(1)$. By Theorem 1, the Donsker property of the class $\{\phi_{jl}^{\Lambda}(t;{\boldsymbol{z}}_0):t\in [0,\tau]\}$, and arguments similar to those used in the proof of proposition 7.27 in \citet{Kosorok08}, $B_{n,l}^{\prime\prime\prime\prime}=o_p(1)$. For $C_{n,l}^{\prime\prime\prime\prime}$, the integrand converges uniformly to 0 in probability and, thus, by Theorem 3 and arguments similar to those used in the proof of proposition 7.27 in \citet{Kosorok08}, it follows that $C_{n,l}^{\prime\prime\prime\prime}=o_p(1)$. Using similar arguments, it can be shown that $D_{n,l}^{\prime\prime\prime\prime}=o_p(1)$. For $E_{n,l}^{\prime\prime\prime\prime}$, the integrand converges uniformly to 0 in probability and, thus, by condition C1 it follows that $E_{n,l}^{\prime\prime\prime\prime}=o_p(1)$. Finally, by Theorem 1, condition C1, and the central limit theorem, it follows that $F_{n,l}^{\prime\prime\prime\prime}=o_p(1)$. Therefore, 
\[
\left\|\hat{W}_{n, l}^{F}(t ; \boldsymbol{z}_{0})-\tilde{W}_{n, l}^{F}(t ; \boldsymbol{z}_{0})\right\|_{\infty}=o_{p}(1),
\]
which completes the proof of the last statement in Theorem 4.

\subsection{Standard Error Estimators}
\label{s:webB5}

The covariance matrix $\boldsymbol{\Sigma}_{l}$ can be consistently estimated using the empirical versions of the influence functions by
\[
\hat{\boldsymbol{\Sigma}}_{l}=\frac{1}{n} \sum_{i=1}^{n} \left\{\frac{1}{M_{i}} \sum_{j=1}^{M_{i}}(\hat{\boldsymbol{\psi}}_{i j l}+\hat{\boldsymbol{R}}_{l}\hat{\boldsymbol{\omega}}_{i j})\right\}^{\otimes 2}.
\]
The first component is
\[
\hat{\boldsymbol{\psi}}_{i j l}={\boldsymbol{H}}_{n,l}^{-1}(\hat{\boldsymbol{\beta}}_{n,l})\int_0^{\tau}\{{\boldsymbol{Z}}_{i j}-\boldsymbol{E}_{n}(t,\hat{\boldsymbol{\beta}}_{n,l})\}d\hat{M}_{i j l}(t;\hat{\boldsymbol{\beta}}_{n,l},\hat{\boldsymbol{\gamma}}_{n}), 
\]
where
\[
\boldsymbol{H}_{n, l}(\hat{\boldsymbol{\beta}}_{n,l}; \hat{\boldsymbol{\gamma}}_n )=\frac{1}{n} \sum_{i=1}^{n} \frac{1}{M_{i}} \sum_{j=1}^{M_{i}} \int_{0}^{\tau} \boldsymbol{V}_{n, l}(t, \hat{\boldsymbol{\beta}}_{n,l}) d \tilde{N}_{i j l}(t ; \hat{\boldsymbol{\gamma}}_n),
\]
\[
\boldsymbol{V}_{n, l}(t, \hat{\boldsymbol{\beta}}_{n,l})=\frac{\sum_{p=1}^{n} \frac{1}{M_{p}} \sum_{q=1}^{M_{p}} Y_{p q}(t) \exp (\hat{\boldsymbol{\beta}}_{n,l}^{T} \boldsymbol{Z}_{p q}) \boldsymbol{Z}_{p q}^{\otimes 2}}{\sum_{p=1}^{n} \frac{1}{M_{p}} \sum_{q=1}^{M_{p}} Y_{p q}(t) \exp (\hat{\boldsymbol{\beta}}_{n,l}^{T} \boldsymbol{Z}_{p q})}-\left\{\frac{\sum_{p=1}^{n} \frac{1}{M_{p}} \sum_{q=1}^{M_{p}} Y_{p q}(t) \exp (\hat{\boldsymbol{\beta}}_{n,l}^{T} \boldsymbol{Z}_{p q}) \boldsymbol{Z}_{p q}}{\sum_{p=1}^{n} \frac{1}{M_{p}} \sum_{q=1}^{M_{p}} Y_{p q}(t) \exp (\hat{\boldsymbol{\beta}}_{n,l}^{T} \boldsymbol{Z}_{p q})}\right\}^{\otimes 2},
\]
\[
\boldsymbol{E}_n(t, \hat{\boldsymbol{\beta}}_{n,l})=\frac{\sum_{p=1}^{n} \frac{1}{M_{p}} \sum_{q=1}^{M_{p}} Y_{p q}(t) \exp (\hat{\boldsymbol{\beta}}_{n,l}^{T} \boldsymbol{Z}_{p q}) \boldsymbol{Z}_{p q}}{\sum_{p=1}^{n} \frac{1}{M_{p}} \sum_{q=1}^{M_{p}} Y_{p q}(t) \exp (\hat{\boldsymbol{\beta}}_{n,l}^{T} \boldsymbol{Z}_{p q})},
\]
and 
\[
\hat{M}_{i j l}(t;\hat{\boldsymbol{\beta}}_{n,l},\hat{\boldsymbol{\gamma}}_{n})=\tilde{N}_{i j l}(t ; \hat{\boldsymbol{\gamma}}_{n})-\int_{0}^{t} Y_{i j}(u) \exp (\hat{\boldsymbol{\beta}}_{n, l}^{T} \boldsymbol{Z}_{i j}) d \hat{\Lambda}_{n, l}(u).
\]

The second component is
\[
\hat{\boldsymbol{R}}_{l}={\boldsymbol{H}}_{n,l}^{-1}(\hat{\boldsymbol{\beta}}_{n,l}) \frac{1}{n} \sum_{i=1}^{n}\left[\frac{1}{M_i} \sum_{j=1}^{M_i} (1-R_{i j})\int_{0}^{\tau}\{\boldsymbol{Z}_{i j}-\boldsymbol{E}_n(t, \hat{\boldsymbol{\beta}}_{n, l})\} d N_{i j}(t)\dot{\pi}_{l}(\boldsymbol{W}_{i j}, \hat{\boldsymbol{\gamma}}_{n})^{T}\right].
\]
A consistent estimator for the covariance function for the baseline cumulative cause-specific hazard is
\[
\frac{1}{n} \sum_{i=1}^{n}\left[\frac{1}{M_{i}} \sum_{j=1}^{M_{i}}\left\{\hat{\phi}_{i j l}(t)+\hat{\boldsymbol{R}}_{l}^{*}(t) \hat{\boldsymbol{\omega}}_{i j}\right\}\right]\left[\frac{1}{M_{i}} \sum_{j=1}^{M_{i}}\left\{\hat{\phi}_{i j l}(s)+\hat{\boldsymbol{R}}_{l}^{*}(s) \hat{\boldsymbol{\omega}}_{i j}\right\}\right], \quad\ t,s \in [0, \tau],
\]
where
\[
\hat{\phi}_{i j l}(t)=\int_0^t\frac{d\hat{M}_{i j l}(t;\hat{\boldsymbol{\beta}}_{n,l},\hat{\boldsymbol{\gamma}}_{n})}{\frac{1}{n} \sum_{i=1}^{n}\{\frac{1}{M_i} \sum_{j=1}^{M_i} Y_{i j}(s) \exp (\hat{\boldsymbol{\beta}}_{n,l}^{T} \boldsymbol{Z}_{i j})\}}-(\hat{\boldsymbol{\psi}}_{ijl}+{\hat{\boldsymbol{R}}}_l\hat{\boldsymbol{\omega}}_{ij})^T\int_0^t \boldsymbol{E}_n(s,\hat{\boldsymbol{\beta}}_{n,l})d\hat{\Lambda}_{n,l}(s),
\]
and
\[
\hat{\boldsymbol{R}}_{l}^{*}(t)=E\left[\frac{1}{M} \sum_{j=1}^{M} (1-R_{j})\dot{\pi}_j({\boldsymbol{W}_{j}},\boldsymbol{\gamma}_0)\int_0^t\frac{dN_{j}(s)}{E\{\frac{1}{M} \sum_{j=1}^{M} Y_{j}(s) \exp (\boldsymbol{\beta}_{0,l}^{T} \boldsymbol{Z}_{j})\}}\right]^T.
\]
A consistent estimator for the covariance function for the covariate-specific cumulative incidence function is
\[
\frac{1}{n} \sum_{i=1}^{n}\left\{ \frac{1}{M_{i}} \sum_{j=1}^{M_{i}} \hat{\phi}_{i j l}^{F}(t ; \boldsymbol{z}_{0})\right\}\left\{ \frac{1}{M_{i}} \sum_{j=1}^{M_{i}} \hat{\phi}_{i j l}^{F}(s; \boldsymbol{z}_{0})\right\}, \quad\ t, s \in[0, \tau],
\]
where
\begin{eqnarray*}
& &\hat{\phi}_{ijl}^F(t;{\boldsymbol{z}}_0)=\int_0^t\exp\left\{-\sum_{l=1}^k\hat{\Lambda}_{n,l}(s-;{\boldsymbol{z}}_0)\right\}d\hat{\phi}^{\Lambda}_{ijl}(s;{\boldsymbol{z}}_0)\\
& &\hspace{25mm}-\int_0^t\left\{\sum_{l=1}^k\hat{\phi}^{\Lambda}_{ijl}(s-;{\boldsymbol{z}}_0)\right\}\exp\left\{-\sum_{l=1}^k\hat{\Lambda}_{n,l}(s-;{\boldsymbol{z}}_0)\right\}d\hat{\Lambda}_{n,l}(s;{\boldsymbol{z}}_0),
\end{eqnarray*}
and
\[
\hat{\phi}_{ijl}^{\Lambda}(t;{\boldsymbol{z}}_0)=\{{\boldsymbol{z}}_0^T(\hat{\boldsymbol{\psi}}_{ijl}+{\hat{\boldsymbol{R}}}_l\hat{\boldsymbol{\omega}}_{ij})\hat{\Lambda}_{n,l}(t)+\hat{\phi}_{ijl}(t)+{\hat{\boldsymbol{R}}}_l^{\star}(t)\hat{\boldsymbol{\omega}}_{ij}\}\exp(\hat{\boldsymbol{\beta}}_{n,l}^T{\boldsymbol{z}}_0).
\]

\section{Additional Simulation Results}
\label{s:webC}

The simulation results for the pointwise estimates of the infinite-dimensional parameters $\Lambda_{0,1}(t)$ and $F_{0,1}(t)$ under scenario 1 are provided in Table~\ref{t:Afour} and~\ref{t:Afive}. Simulation results under scenario 2 are presented in Tables~\ref{t:Aone},~\ref{t:Atwo1},~\ref{t:Atwo2} and~\ref{t:Athree}.

\begin{table}
\caption{Simulation results for the infinite-dimensional parameter $\Lambda_{0,1}(t)$ at selected time points under Scenario 1. Results from the proposed approach and the approach by \citet{Bakoyannis20} (BZY20) which ignores the within-cluster dependence.}
\label{t:Afour}
\begin{center}
\begin{tabular}{lllcccccccc}
\hline
& & & \multicolumn{4}{c}{Proposed} & \multicolumn{4}{c}{BZY20}\\
\cmidrule(lr){4-7}\cmidrule(lr){8-11}
$n$ & $p_m(\%)$ & $t$ & Bias & MCSD & ASE & CP & Bias & MCSD & ASE & CP \\
\hline
50	&25	&0.1	&0.001	&0.084	&0.080	&0.915	&0.090	&0.104	&0.025	&0.252\\
	&	&0.2	&0.000	&0.105	&0.101	&0.917	&0.120	&0.127	&0.034	&0.259\\
	&	&0.4	&0.004	&0.135	&0.130	&0.934	&0.160	&0.156	&0.049	&0.284\\
	&	&0.8	&0.011	&0.185	&0.176	&0.947	&0.203	&0.201	&0.081	&0.385\\
	&43	&0.1	&0.000	&0.086	&0.081	&0.915	&0.091	&0.106	&0.031	&0.306\\
	&	&0.2	&-0.001	&0.108	&0.104	&0.926	&0.121	&0.130	&0.042	&0.318\\
	&	&0.4	&0.003	&0.139	&0.134	&0.935	&0.160	&0.160	&0.058	&0.355\\
	&	&0.8	&0.011	&0.188	&0.181	&0.942	&0.203	&0.203	&0.093	&0.441\\
200	&25	&0.1	&0.001	&0.040	&0.040	&0.954	&0.090	&0.049	&0.013	&0.074\\
	&	&0.2	&0.001	&0.051	&0.051	&0.951	&0.122	&0.062	&0.017	&0.055\\
	&	&0.4	&0.003	&0.066	&0.066	&0.955	&0.159	&0.077	&0.024	&0.055\\
	&	&0.8	&0.006	&0.093	&0.088	&0.950	&0.200	&0.101	&0.040	&0.096\\
	&43	&0.1	&0.001	&0.040	&0.041	&0.952	&0.091	&0.051	&0.015	&0.096\\
	&	&0.2	&0.002	&0.053	&0.052	&0.952	&0.123	&0.064	&0.021	&0.081\\
	&	&0.4	&0.003	&0.069	&0.068	&0.952	&0.161	&0.079	&0.029	&0.077\\
	&	&0.8	&0.007	&0.096	&0.091	&0.947	&0.199	&0.103	&0.046	&0.138\\
\hline
\end{tabular}
\end{center}
{\footnotesize Note: $n$: number of clusters with cluster size $M \in [30,60]$; $p_m$: percentage of missingness; $t$: selected time points ; MCSD: Monte Carlo standard deviation; ASE: average estimated standard error; CP: coverage probability of $95\%$ pointwise confidence interval}
\end{table}

\begin{table}
\caption{Simulation results for the infinite-dimensional parameter $F_{0,1}(t)$ at selected time points under Scenario 1. Results from the proposed approach and the approach by \citet{Bakoyannis20} (BZY20) which ignores the within-cluster dependence.}
\label{t:Afive}
\begin{center}
\begin{tabular}{lllcccccccc}
\hline
& & & \multicolumn{4}{c}{Proposed} & \multicolumn{4}{c}{BZY20}\\
\cmidrule(lr){4-7}\cmidrule(lr){8-11}
$n$ & $p_m(\%)$ & $t$ & Bias & MCSD & ASE & CP & Bias & MCSD & ASE & CP \\
\hline
50	&25	&0.1	&-0.002	&0.049	&0.047	&0.924	&0.036	&0.055	&0.014	&0.300\\
	&	&0.2	&-0.003	&0.051	&0.049	&0.926	&0.033	&0.057	&0.015	&0.343\\
	&	&0.4	&-0.002	&0.053	&0.051	&0.933	&0.025	&0.058	&0.017	&0.388\\
	&	&0.8	&-0.002	&0.053	&0.051	&0.938	&0.015	&0.058	&0.018	&0.429\\
	&43	&0.1	&-0.002	&0.049	&0.047	&0.924	&0.036	&0.056	&0.018	&0.381\\
	&	&0.2	&-0.003	&0.053	&0.051	&0.924	&0.033	&0.059	&0.020	&0.427\\
	&	&0.4	&-0.002	&0.055	&0.053	&0.934	&0.026	&0.060	&0.023	&0.493\\
	&	&0.8	&-0.002	&0.056	&0.054	&0.936	&0.015	&0.060	&0.024	&0.555\\
200	&25	&0.1	&0.000	&0.023	&0.024	&0.954	&0.038	&0.027	&0.007	&0.177\\
	&	&0.2	&0.000	&0.025	&0.025	&0.950	&0.035	&0.028	&0.008	&0.198\\
	&	&0.4	&0.000	&0.026	&0.026	&0.952	&0.028	&0.029	&0.008	&0.287\\
	&	&0.8	&0.000	&0.027	&0.026	&0.945	&0.016	&0.029	&0.009	&0.402\\
	&43	&0.1	&0.000	&0.024	&0.024	&0.952	&0.038	&0.028	&0.009	&0.204\\
	&	&0.2	&0.000	&0.026	&0.026	&0.953	&0.036	&0.029	&0.010	&0.276\\
	&	&0.4	&0.000	&0.028	&0.027	&0.949	&0.028	&0.030	&0.011	&0.368\\
	&	&0.8	&0.000	&0.028	&0.027	&0.945	&0.017	&0.030	&0.012	&0.503\\
\hline
\end{tabular}
\end{center}
{\footnotesize Note: $n$: number of clusters with cluster size $M \in [30,60]$; $p_m$: percentage of missingness; $t$: selected time points ; MCSD: Monte Carlo standard deviation; ASE: average estimated standard error; CP: coverage probability of $95\%$ pointwise confidence interval}
\end{table}

\begin{table}
\caption{Simulation results for the regression coefficient $\beta_{1}$ under Scenario 2 for the proposed approach and the approach by \citet{Bakoyannis20} (BZY20) which ignores the within-cluster dependence.}
\label{t:Aone}
\begin{center}
\begin{tabular}{llcccccccc}
\hline
& & \multicolumn{4}{c}{Proposed} & \multicolumn{4}{c}{BZY20}\\
\cmidrule(lr){3-6}\cmidrule(lr){7-10}
$n$ & $p_m(\%)$ & Bias & MCSD & ASE & CP & Bias & MCSD & ASE & CP \\
\hline
50	&26	&-0.002	&0.035	&0.033	&0.938	&0.008	&0.035	&0.023	&0.781\\
	&36	&0.000	&0.036	&0.035	&0.936	&0.011	&0.036	&0.025	&0.803\\
	&44	&0.002	&0.037	&0.036	&0.936	&0.013	&0.037	&0.027	&0.832\\
100	&26	&0.002	&0.024	&0.024	&0.950	&0.012	&0.023	&0.016	&0.740\\
	&36	&0.004	&0.025	&0.025	&0.937	&0.014	&0.024	&0.018	&0.771\\
	&44	&0.005	&0.026	&0.026	&0.937	&0.016	&0.025	&0.019	&0.782\\
200	&26	&0.003	&0.017	&0.017	&0.945	&0.013	&0.016	&0.011	&0.681\\
	&36	&0.005	&0.017	&0.018	&0.943	&0.016	&0.017	&0.012	&0.672\\
	&44	&0.007	&0.018	&0.018	&0.940	&0.018	&0.018	&0.014	&0.663\\
\hline\\
\end{tabular}
\end{center}
{\footnotesize Note: $n$: number of clusters with cluster size $M \in [30,60]$; $p_m$: percentage of missingness; MCSD: Monte Carlo standard deviation; ASE: average estimated standard error; CP: coverage probability of $95\%$ confidence interval}
\end{table}

\begin{table}
\caption{Simulation results for the infinite-dimensional parameter $\Lambda_{0,1}(t)$ at selected time points under Scenario 2. Results from the proposed approach and the approach by \citet{Bakoyannis20} (BZY20) which ignores the within-cluster dependence.}
\label{t:Atwo1}
\begin{center}
\begin{tabular}{lllcccccccc}
\hline
& & & \multicolumn{4}{c}{Proposed} & \multicolumn{4}{c}{BZY20}\\
\cmidrule(lr){4-7}\cmidrule(lr){8-11}
$n$ & $p_m(\%)$ & $t$ & Bias & MCSD & ASE & CP & Bias & MCSD & ASE & CP \\
\hline
50	&26	&0.1	&-0.004	&0.088	&0.083	&0.911	&0.084	&0.106	&0.028	&0.308\\
	&	&0.2	&-0.013	&0.109	&0.104	&0.907	&0.104	&0.129	&0.037	&0.317\\
	&	&0.4	&-0.014	&0.141	&0.134	&0.922	&0.139	&0.161	&0.052	&0.339\\
	&	&0.8	&0.004	&0.193	&0.182	&0.939	&0.200	&0.210	&0.083	&0.405\\
	&44	&0.1	&-0.008	&0.088	&0.083	&0.907	&0.078	&0.106	&0.034	&0.376\\
	&	&0.2	&-0.024	&0.110	&0.105	&0.897	&0.090	&0.130	&0.044	&0.403\\
	&	&0.4	&-0.028	&0.142	&0.136	&0.911	&0.121	&0.162	&0.061	&0.422\\
	&	&0.8	&-0.001	&0.196	&0.186	&0.945	&0.194	&0.214	&0.095	&0.461\\
200	&26	&0.1	&-0.004	&0.041	&0.042	&0.955	&0.085	&0.050	&0.014	&0.101\\
	&	&0.2	&-0.012	&0.053	&0.052	&0.931	&0.106	&0.062	&0.018	&0.109\\
	&	&0.4	&-0.015	&0.068	&0.067	&0.934	&0.139	&0.078	&0.026	&0.111\\
	&	&0.8	&-0.001	&0.096	&0.091	&0.948	&0.196	&0.104	&0.041	&0.111\\
	&44	&0.1	&-0.007	&0.041	&0.042	&0.942	&0.079	&0.050	&0.017	&0.163\\
	&	&0.2	&-0.021	&0.053	&0.053	&0.916	&0.092	&0.063	&0.022	&0.201\\
	&	&0.4	&-0.029	&0.069	&0.068	&0.916	&0.121	&0.079	&0.030	&0.205\\
	&	&0.8	&-0.006	&0.098	&0.093	&0.940	&0.188	&0.106	&0.047	&0.168\\
\hline
\end{tabular}
\end{center}
{\footnotesize Note: $n$: number of clusters with cluster size $M \in [30,60]$; $p_m$: percentage of missingness; $t$: selected time points ; MCSD: Monte Carlo standard deviation; ASE: average estimated standard error; CP: coverage probability of $95\%$ pointwise confidence interval}
\end{table}

\begin{table}
\caption{Simulation results for the infinite-dimensional parameters $F_{0,1}(t)$ at selected time points under Scenario 2. Results from the proposed approach and the approach by \citet{Bakoyannis20} (BZY20) which ignores the within-cluster dependence.}
\label{t:Atwo2}
\begin{center}
\begin{tabular}{lllcccccccc}
\hline
& & & \multicolumn{4}{c}{Proposed} & \multicolumn{4}{c}{BZY20}\\
\cmidrule(lr){4-7}\cmidrule(lr){8-11}
$n$ & $p_m(\%)$ & $t$ & Bias & MCSD & ASE & CP & Bias & MCSD & ASE & CP \\
\hline
50	&26	&0.1	&0.001	&0.045	&0.043	&0.928	&0.024	&0.051	&0.013	&0.368\\
	&	&0.2	&-0.002	&0.048	&0.046	&0.925	&0.017	&0.054	&0.015	&0.409\\
	&	&0.4	&-0.003	&0.051	&0.049	&0.931	&0.008	&0.056	&0.016	&0.425\\
	&	&0.8	&-0.001	&0.052	&0.050	&0.932	&-0.002	&0.057	&0.017	&0.448\\
	&44	&0.1	&0.004	&0.046	&0.044	&0.929	&0.026	&0.052	&0.018	&0.452\\
	&	&0.2	&-0.002	&0.050	&0.048	&0.922	&0.017	&0.055	&0.020	&0.483\\
	&	&0.4	&-0.004	&0.053	&0.051	&0.927	&0.008	&0.058	&0.022	&0.533\\
	&	&0.8	&-0.001	&0.054	&0.052	&0.933	&-0.001	&0.058	&0.023	&0.567\\
200	&26	&0.1	&0.003	&0.022	&0.022	&0.957	&0.025	&0.025	&0.007	&0.270\\
	&	&0.2	&0.000	&0.024	&0.024	&0.947	&0.019	&0.027	&0.007	&0.347\\
	&	&0.4	&-0.002	&0.025	&0.025	&0.943	&0.009	&0.028	&0.008	&0.420\\
	&	&0.8	&-0.001	&0.026	&0.025	&0.944	&-0.001	&0.029	&0.009	&0.442\\
	&44	&0.1	&0.006	&0.022	&0.023	&0.952	&0.029	&0.026	&0.009	&0.306\\
	&	&0.2	&0.000	&0.025	&0.024	&0.942	&0.019	&0.028	&0.010	&0.434\\
	&	&0.4	&-0.002	&0.026	&0.026	&0.945	&0.009	&0.029	&0.011	&0.517\\
	&	&0.8	&0.000	&0.027	&0.027	&0.939	&0.000	&0.030	&0.012	&0.554\\
\hline
\end{tabular}
\end{center}
{\footnotesize Note: $n$: number of clusters with cluster size $M \in [30,60]$; $p_m$: percentage of missingness; $t$: selected time points ; MCSD: Monte Carlo standard deviation; ASE: average estimated standard error; CP: coverage probability of $95\%$ pointwise confidence interval}
\end{table}

\begin{table}
\caption{Simulation results for the coverage probabilities of $95\%$ simultaneous confidence bands for the infinite-dimensional parameters $\Lambda_{0,1}(t)$ and $F_{0,1}(t)$ under Scenario 2. Results from the proposed approach and the approach by \citet{Bakoyannis20} (BZY20) which ignores the within-cluster dependence.}
\label{t:Athree}
\begin{center}
\begin{tabular}{llcccccccc}
\hline
& & \multicolumn{4}{c}{$\Lambda_{0,1}(t)$} & \multicolumn{4}{c}{$F_{0,1}(t)$}\\
\cmidrule(lr){3-6}\cmidrule(lr){7-10}
$n$ & $p_m(\%)$ & \multicolumn{2}{c}{Proposed} & \multicolumn{2}{c}{BZY20} & \multicolumn{2}{c}{Proposed} & \multicolumn{2}{c}{BZY20} \\
\cmidrule(lr){3-4}\cmidrule(lr){5-6}\cmidrule(lr){7-8}\cmidrule(lr){9-10}			
& & EP & HW & EP & HW & EP & HW & EP & HW \\
\hline
50	&26	&0.768	&0.914	&0.078	&0.152	&0.803	&0.924	&0.087	&0.171\\
	&36	&0.684	&0.908	&0.073	&0.166	&0.731	&0.923	&0.081	&0.169\\
	&44	&0.598	&0.915	&0.078	&0.163	&0.661	&0.927	&0.078	&0.172\\
100	&26	&0.755	&0.942	&0.038	&0.098	&0.796	&0.935	&0.035	&0.113\\
	&36	&0.568	&0.938	&0.026	&0.089	&0.609	&0.931	&0.024	&0.089\\
	&44	&0.421	&0.941	&0.013	&0.069	&0.452	&0.927	&0.010	&0.060\\
200	&26	&0.684	&0.946	&0.009	&0.045	&0.717	&0.940	&0.013	&0.065\\
	&36	&0.348	&0.942	&0.002	&0.035	&0.358	&0.929	&0.001	&0.034\\
	&44	&0.169	&0.922	&0.000	&0.012	&0.179	&0.914	&0.000	&0.013\\
\hline\\
\end{tabular}
\end{center}
{\footnotesize Note: $n$: number of clusters with cluster size $M \in [30,60]$; $p_m$: percentage of missingness; EP: equal precision bands; HW: Hall-Wellner-type bands}
\end{table}

\end{document}